\documentclass[referee]{aa} 
\usepackage{graphicx}
\usepackage{txfonts}
\newcounter{newctr}
\begin{document}
 \title{The high-velocity clouds and the Magellanic Clouds}

 \author{C. A. Olano
 \thanks{Member of the Carrera del Investigador Cient\'{\i}fico of the 
CONICET, Argentina} }

 \offprints{C. A. Olano}

 \institute{Universidad Nacional de La Plata, Facultad de Ciencias Astron\'{o}micas y Geof\'{\i}sicas, Paseo del Bosque, 1900 La Plata, Argentina\\
\email{colano@lilen.fcaglp.unlp.edu.ar}}

 \date{Received date ; Accepted date}

\abstract{  From an analysis of  the sky and velocity distributions of the high-velocity clouds (HVCs) 
we show that the majority of the HVCs has a common origin. We conclude that the HVCs surround the Galaxy, forming a metacloud of $\sim 300 \mbox{ kpc}$
in size and with a mass of $\sim 3\times 10^{9}\, \mathrm{ M}_{\sun}$,  and that they are the product of a powerful ``superwind'' (about $10^{58} \mbox{ ergs}$), which  occurred in the Magellanic Clouds about $570 \mbox{ Myr}$ ago
as a consequence of the interaction of the Large and Small  Magellanic Clouds. The HVCs might be magnetic bubbles of semi-ionized gas, blown  from  the Magellanic Clouds  around $570 \mbox{ Myr}$ ago, that  circulate  largely through the halo of the Galaxy as a stream or flow of gas.

 On the basis of the connection  found
between the HVCs and the Magellanic Clouds, we have constructed a theoretical model with the purpose of computing the orbits of a sample of test particles representing the HVCs, under the gravitational action  of the Galaxy and the Magellanic Clouds.
The orbits of the Large  and Small Magellanic Clouds have been traced backwards in time to estimate  the position and velocity of the Clouds at the time of the collision between the two Clouds, and to infer  the initial conditions of the HVCs.  The model can reproduce   the main features of position and velocity distributions of the HVCs,  like the overall structure and kinematics of  the Magellanic Stream. The initial  velocities of the HVCs were the result of velocities of  expansion that permitted the escape of the HVCs from the Magellanic Clouds   plus  the systemic velocity of the Magellanic Clouds at the time of the collision. With these initial conditions,  the Galactic gravitational potential  induced   differential rotations or  shearing motions that  elongated the cloud of  HVCs in the orbital direction,  forming the rear and front parts of the Magellanic stream. The population  of HVCs is centered around the Magellanic Clouds. The eccentric position of the Sun within the cloud  of HVCs explains the asymmetries between   the sky distributions of the HVCs  of the northern   Galactic hemisphere  and those of the   southern  Galactic hemisphere.    

In the light of the  model we analyze the effects that  the passage of the HVC flow through the Galactic disk has produced on the interstellar medium. The effects of the HVC flow  can account for many observational details  such as  the Galactic warp, HI shells and supershells in the gaseous layer of  the outer parts of  the Milky Way.  The Galactic disk was target of numerous impacts of HVCs in the course of  the last $400\,\mbox{Myr}$, accumulating mass at the average rate of approximately $0.6\, \mathrm{M}_{\sun}$  per year. The events  of this period may  be regarded as landmarks in the evolutionary  history of the Milky Way.
  
   }

\maketitle
\titlerunning{The high-velocity clouds and the Magellanic Clouds}
\authorrunning{Olano}
%

\section{Introduction}
The existence of HI high-velocity clouds (HVCs) at high Galactic latitudes has
 intrigued  astronomers  from the time of their discovery in 1963.
The early observations induced the belief that all HVCs had negative 
radial velocities, and that these clouds could be intergalactic gas 
falling onto the Galaxy. The recognition of violent activity in the 
Galactic disk and  the discovery of large complexes of high positive velocity
clouds  and of the Magellanic Stream opened a whole gamut of possibilities for
 the nature
and origin of the HVCs  such as ``Galactic fountain'' clouds, material stripped
from the Magellanic Clouds and  extragalactic objects. The lack of adequate
distance indicators  makes it  difficult to  discriminate among  the different
alternatives. 

The first interpretations of the HVCs were discussed by Oort (\cite{oorta}, \cite{oortb}, \cite{oortc}).   Wakker \& van Woerden (\cite{wakker-woerden}) have given a  recent review  of the HVC phenomenon.
P\"{o}ppel (\cite{poeppel})  reviewed the possible role  played by the HVCs  in the local
interstellar  medium. Although a large amount of observational data  has now been accumulated, such as very 
sensitive HI 21 cm surveys (Morras et al. \cite{morras}; de Heij et al.  \cite{heijA}, \cite{heijB};
 Lockman et al. \cite{lockman};
  Putman et al. \cite{putman}; Putman et al. \cite{putman1}),
 the detection of molecules (Richter et al. \cite{richter}) and H$\mathrm{\alpha}$   emission in HVCs (Tufte et al. \cite{tofte}; Putman et al. \cite{putman2})
and the distance constraints to some HVCs (Wakker \cite{wakker}), as well as  
a large number of theoretical studies that have been dedicated to this topic
 (eg.  Quilis \& Moore \cite{quilis};  Espresate et al. \cite{espresate};  Sternberg et al. \cite{sternberg}; Maloney \& Putman \cite{maloney}), 
the HVCs remain  enigmatic.

The theories for the origin and nature of the high velocity clouds can be divided into three categories: 
 intergalactic (eg. Bajaja et al. \cite{bajaja1}; Blitz et al. \cite{blitz};
 Braun \& Burton \cite{braun}), 
circumgalactic (eg. Kerr \& Sullivan III \cite{kerr}; Hulsbosch \& Oort \cite{hulsbosch}) and 
 Galactic  (eg. Bregman \cite{bregman}; Verschuur \cite{verschuur}).
  On the basis  
of an analysis  of the sky and velocity distributions of the HVCs  we will demonstrate  in this paper  that the majority of these clouds  constitutes  a
 circumgalactic flux or stream of clouds related to  the Magellanic Stream. The idea that the HVCs may be fragments of Magellanic material precipitating toward the galactic disk was first suggested   by  Giovanelli (\cite{giovanelli}) and Mirabel (\cite{mirabela}). The main theories on the origin of 
the Magellanic Stream itself consider  that the stream consists of gas  extracted from
  the Magellanic Clouds. The proposed mechanisms that have been studied in detail are  tidal
 forces of the Galaxy, friction forces of
 the gaseous galactic halo  and  a collision between the Large and Small Magellanic Clouds 
(Gardiner et al. \cite{gardiner2}; Heller \& Rohlfs \cite{heller}; Moore \& Davis \cite{moore}; Gardiner \& Noguchi \cite{gardiner}).
 The tidal forces affect equally both gas
 and  stars of  the Magellanic Clouds. However there are no stars in the Stream, indicating  that 
 tidal interaction could  not be the primary
 cause of its formation. On the other hand, the weakness of the  stripping
 drag hypothesis  is that the
 gas density of the Galactic halo should  be very low. Hence, to account for
 the  formation 
of the Magellanic Stream and  the rest of the HVCs, we will here take a different point of view that includes only
 the collision between the Large and Small Magellanic Clouds as the  mechanism that triggered the process.

 \section{A simple  interpretation of the sky distribution and kinematics of the high velocity clouds}

In general the observations of  HVCs provide  the HVC positions projected on the sky and  their  radial velocities.
 Hence, we do not know   the  distances  and tangential velocities of the HVCs. Even with  this limitation,
  if the HVCs are a  homogeneous population with a common  origin, in principle it is possible to find
 the  space distribution and the three-dimensional velocity distribution  of the HVC population   from the analysis of the observed
 distributions
 of the sky positions and radial velocities of the HVCs. This is a typical case of an inverse problem
 in astronomy (Craig \& Brown \cite{craig}; Merritt \cite{merritt}; Saha \cite{saha}).

 Nowadays a vast amount of observational material is available on HVCs  in the form of surveys
 whose combination  covers  the whole sky,  and that therefore provide   complete and reliable statistical information.
   Wakker (\cite{wakker-privado}) compiled a catalog of HVCs with 11000 entries, using
 the surveys of Hulsbosch \& Wakker (\cite{hulsbosch-wakker}), Bajaja et al. (\cite{bajaja})
 for $-23^{\circ}<\delta <-18^{\circ}$ and
 Morras et al. (\cite{morras}) for $\delta <-23^{\circ}$. Each entry contains the celestial position coordinates, $\ell$  and  $b$,  the radial velocity $\rho$ and other parameters that characterize a detected high-velocity profile component (i.e. a piece of cloud intercepted by the radiotelescope beam). An HVC can be defined by a set of points ($\ell$, $b$, $\rho$), or cloud elements, that satisfy  criteria  of continuity in the ($\ell$, $b$, $\rho$)-space.  For the study of  HVC distributions, we will represent all the HVC components as separate points ($\ell$, $b$, $\rho$); although each point can be associated with a subset of points (defining an HVC or  complex of HVCs)  that are not necessarily independent. In statistical considerations one should take this into account (see Sect. 4). We will represent the distributions of HVCs in the conventional form as orthogonal projections of points in  the ($\ell$, $b$, $\rho$)-space upon the planes $\ell$-$b$, $\ell$-$\rho$  and  $b$-$\rho$, although a three-dimensional representation is  more elegant and compact.

On the basis of the data of Wakker's catalog, we display  the distributions
 of the positions and radial velocities  of the HVCs in Figs.~\ref{DataSky1}, ~\ref{Lon-Vel1} 
and ~\ref{Lat-Vel1}.
\setcounter{newctr}{1}
\renewcommand{\thenewctr}{\alph{newctr}}
\renewcommand{\thefigure}{\arabic{figure}-\thenewctr}
 \begin{figure}
   \centering
 \includegraphics[width=\textwidth]{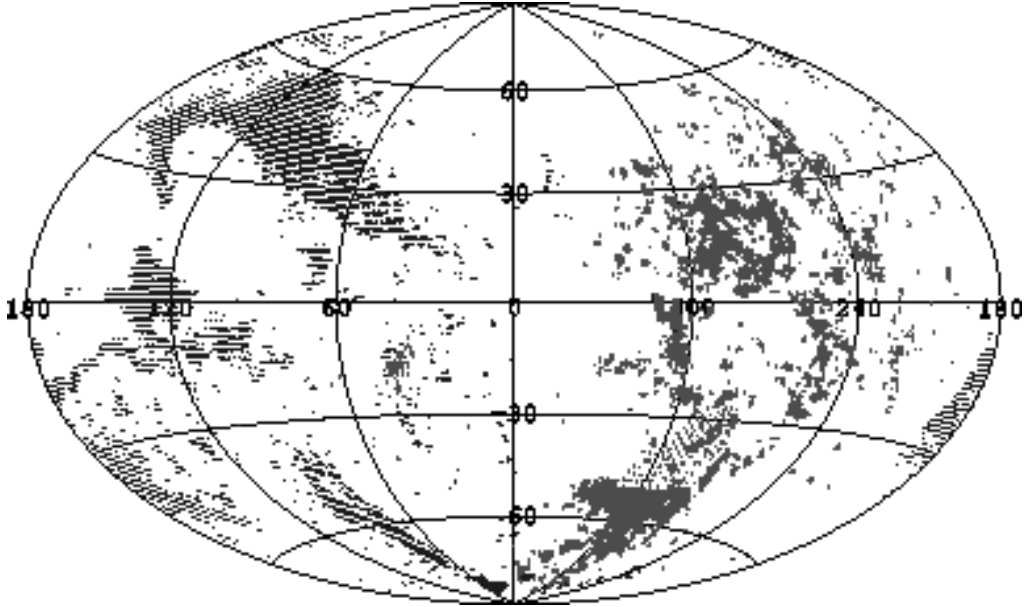}
      \caption{Aitoff projection of the entire  sky distribution of the HVCs in Galactic coordinates, based on measurements of the 21-cm emission line of HI (adapted from  Bajaja et al. \cite{bajaja},  Hulsbosch \& Wakker \cite{hulsbosch-wakker}  and Morras et al. \cite{morras}). Plus (or darker regions) and  minus signs mark respectively the  positions of the  HI lines  with  positive  and negative  high velocities  in the LSR frame.}
         \label{DataSky1}

   \end{figure}
\addtocounter{figure}{-1}
\addtocounter{newctr}{1}
\begin{figure}
   \centering
 \includegraphics[width=\textwidth]{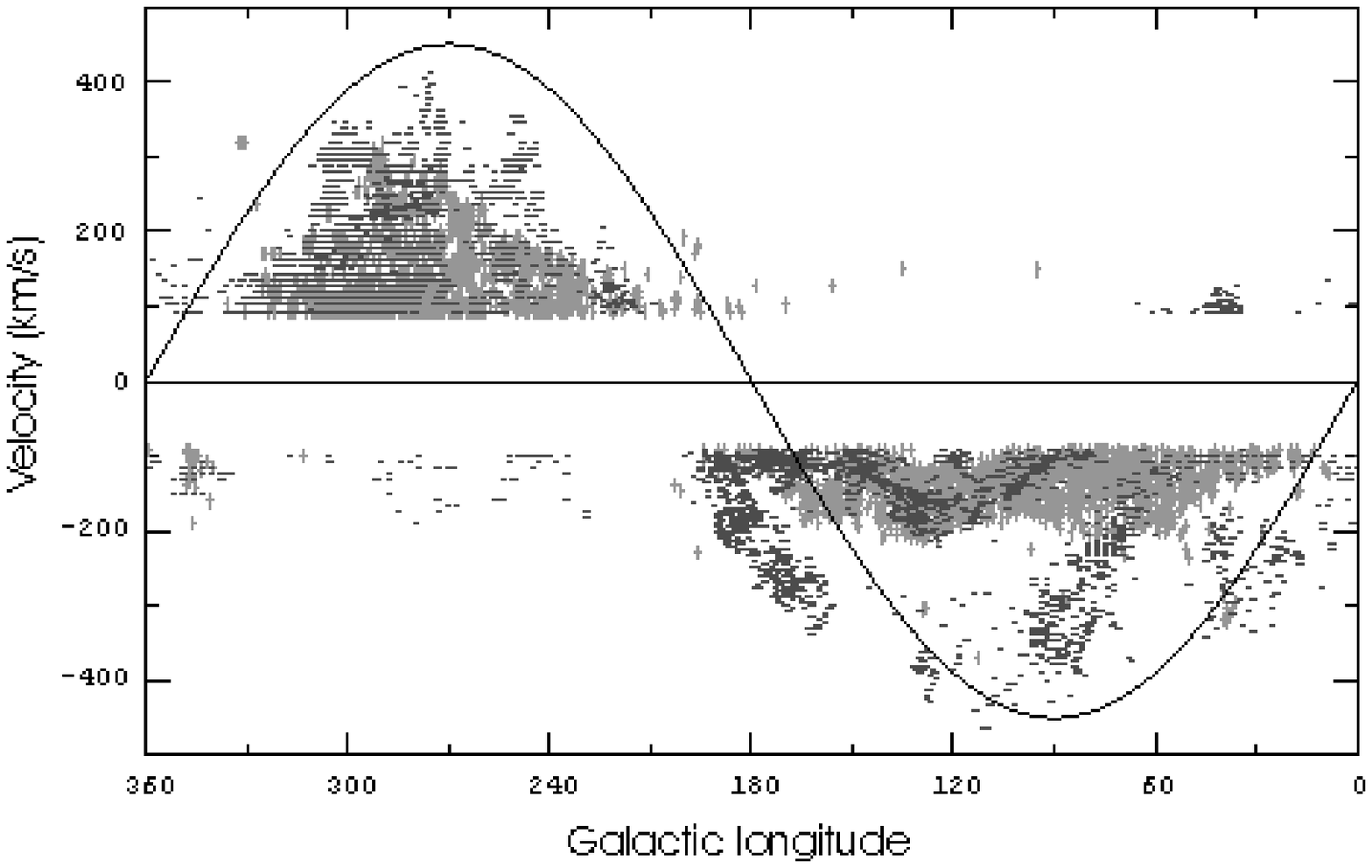}
      \caption{Velocity of high-velocity HI lines  with respect to the LSR plotted against Galactic longitude, from  the  collection of data used in  Fig. ~\ref{DataSky1}. The sinusoidal curve represents  the $V_\mathrm{LSR}(\ell)$ pattern  for  $b=0^{\circ}$ 
corresponding  to  a hypothetical flow of HVCs, circulating in
 directions  parallel to $\ell=90^{\circ}-270^{\circ}$. 
              }
         \label{Lon-Vel1}
   \end{figure}
\addtocounter{figure}{-1}
\addtocounter{newctr}{1}
\begin{figure}
   \centering
 \includegraphics[width=\textwidth]{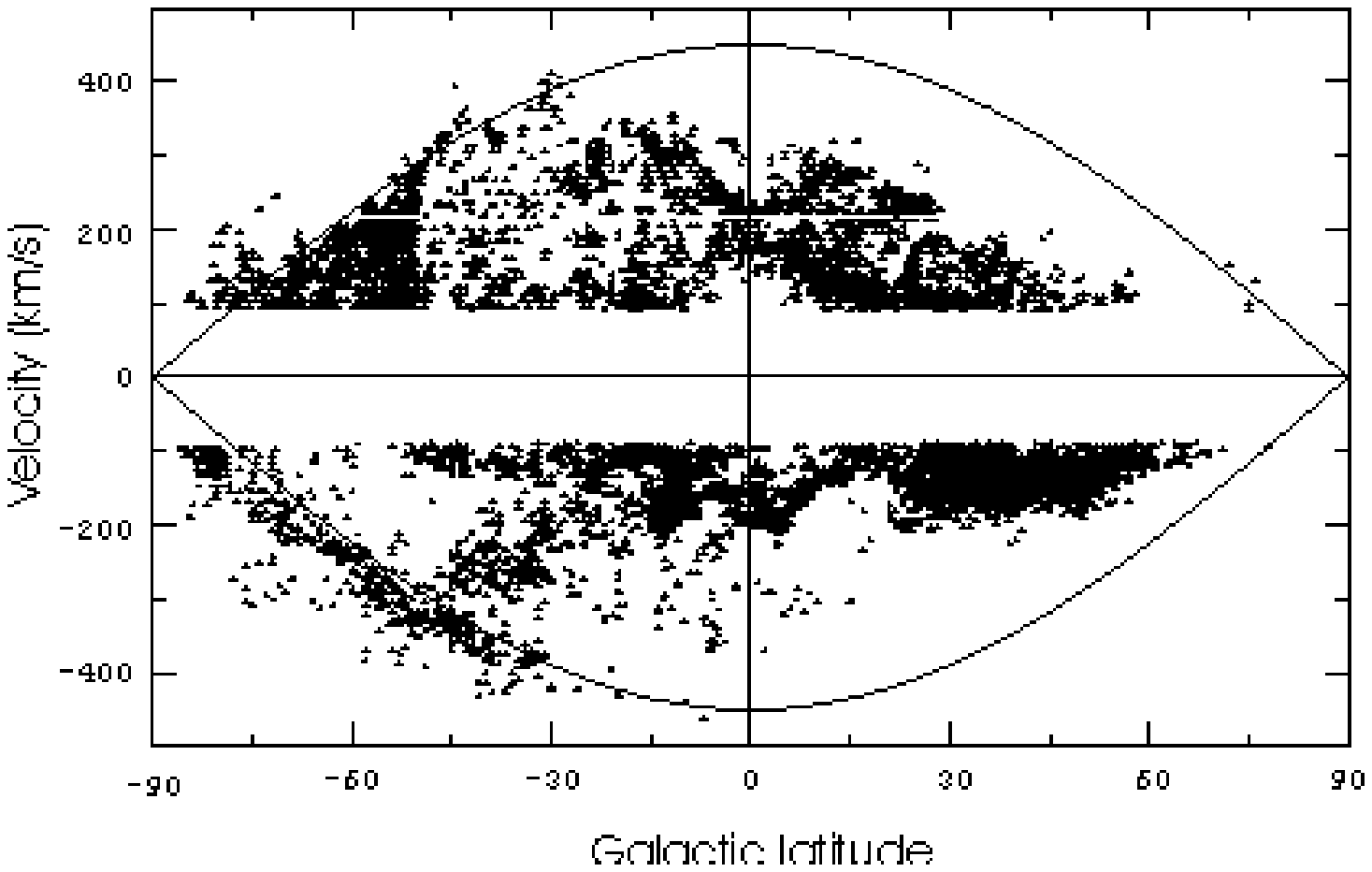}
      \caption{Velocity  of high-velocity HI lines with respect to the LSR plotted against Galactic latitude, from  the  collection of data used in  Fig. ~\ref{DataSky1}. The two half-cosine curves represent  the $V_\mathrm{LSR}(b)$ pattern  for  $\ell=90^{\circ}$ and  $270^{\circ}$ corresponding  to  a hypothetical flow of HVCs, circulating in 
 directions  parallel to $\ell=90^{\circ}-270^{\circ}$. 
              }
         \label{Lat-Vel1}
\end{figure}
For purposes of comparison, in  Figs.~\ref{DataSky2}, ~\ref{Lon-Vel2} 
and ~\ref{Lat-Vel2} we represent  the HVC distributions corresponding to  a second set of data with an improvement 
in spatial resolution  
taken from  Putman et al. (\cite{putman}) and Giovanelli (\cite{giovanelli2}) for $\delta > -15 \degr$.
\setcounter{newctr}{1}  
 \begin{figure}
   \centering
 \includegraphics[width=\textwidth]{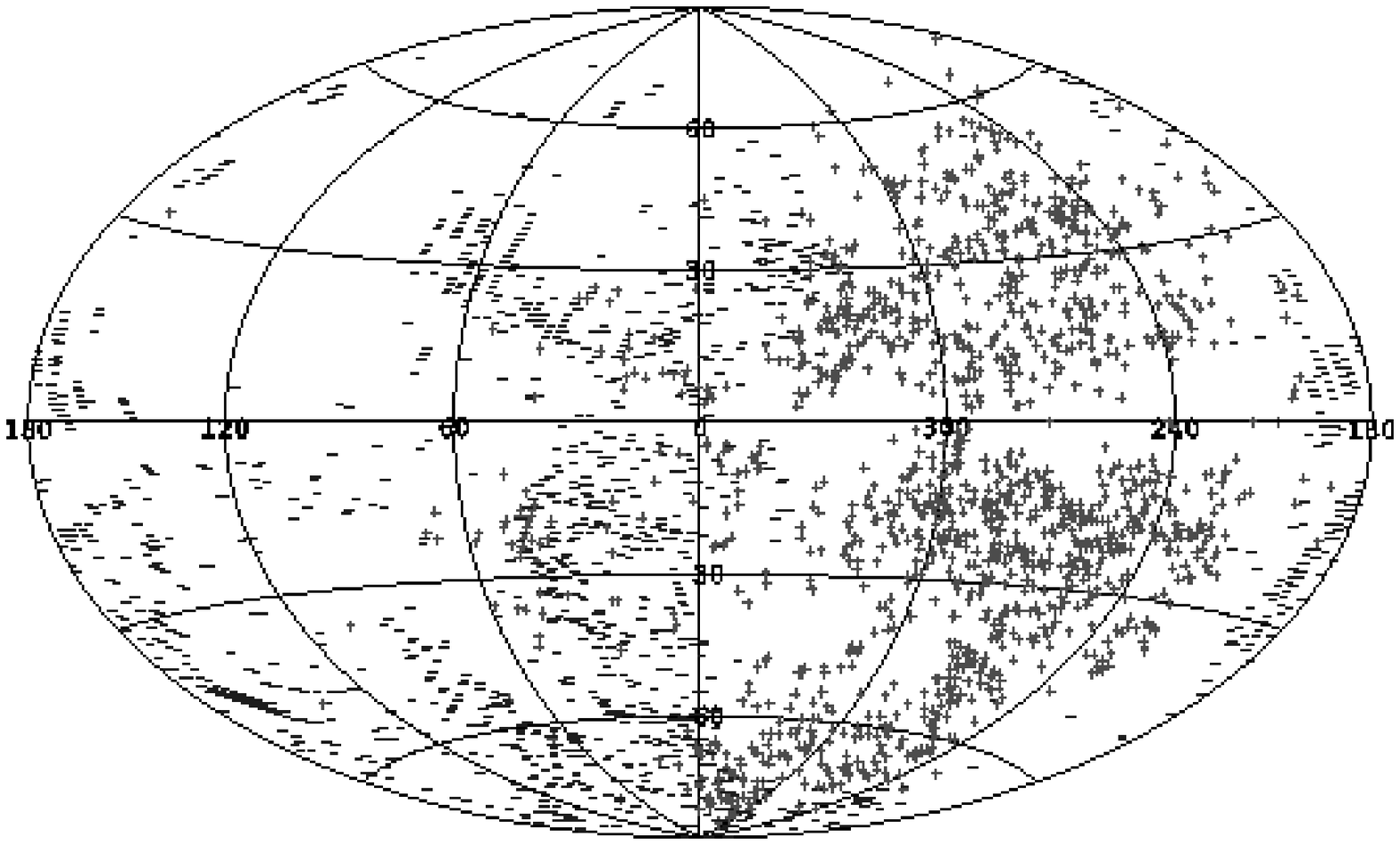}
      \caption{ Same as Fig. ~\ref{DataSky1}, but with data adapted from Giovanelli (\cite{giovanelli}) and Putman (\cite{putman}). 
The region centered at $\ell=120^{\circ}$ and $b=30^{\circ}$ is not covered by the observations of the two surveys used here, and is left empty.    }
         \label{DataSky2}
   \end{figure}

\addtocounter{figure}{-1}
\addtocounter{newctr}{1}

\begin{figure}
   \centering
 \includegraphics[width=\textwidth]{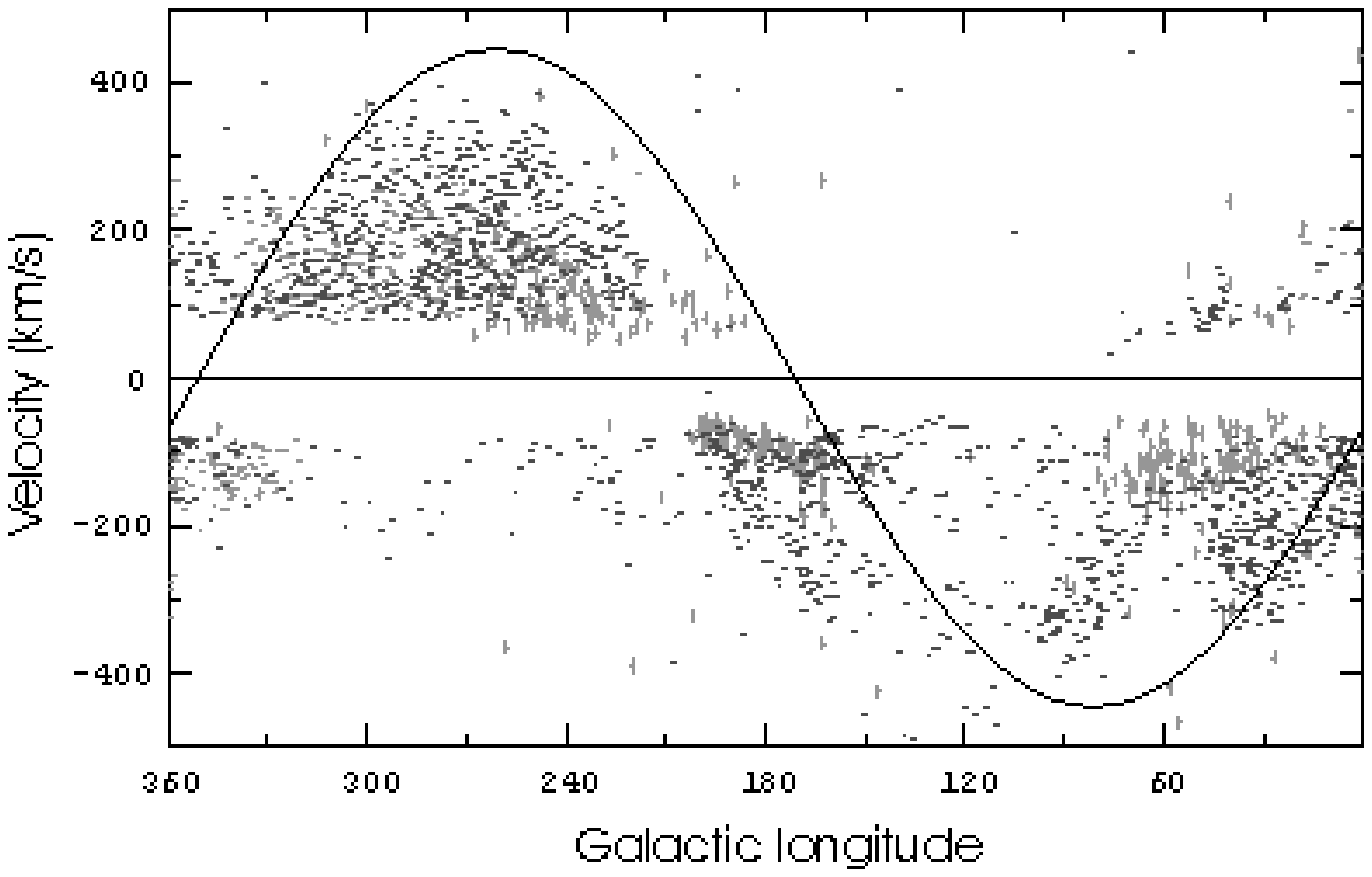}
      \caption{Same as Fig. ~\ref{Lon-Vel1}, but from  the collection of data used in Fig.  ~\ref{DataSky2}.
              }
         \label{Lon-Vel2}
   \end{figure}
\addtocounter{figure}{-1}
\addtocounter{newctr}{1}

\begin{figure}
   \centering
 \includegraphics[width=\textwidth]{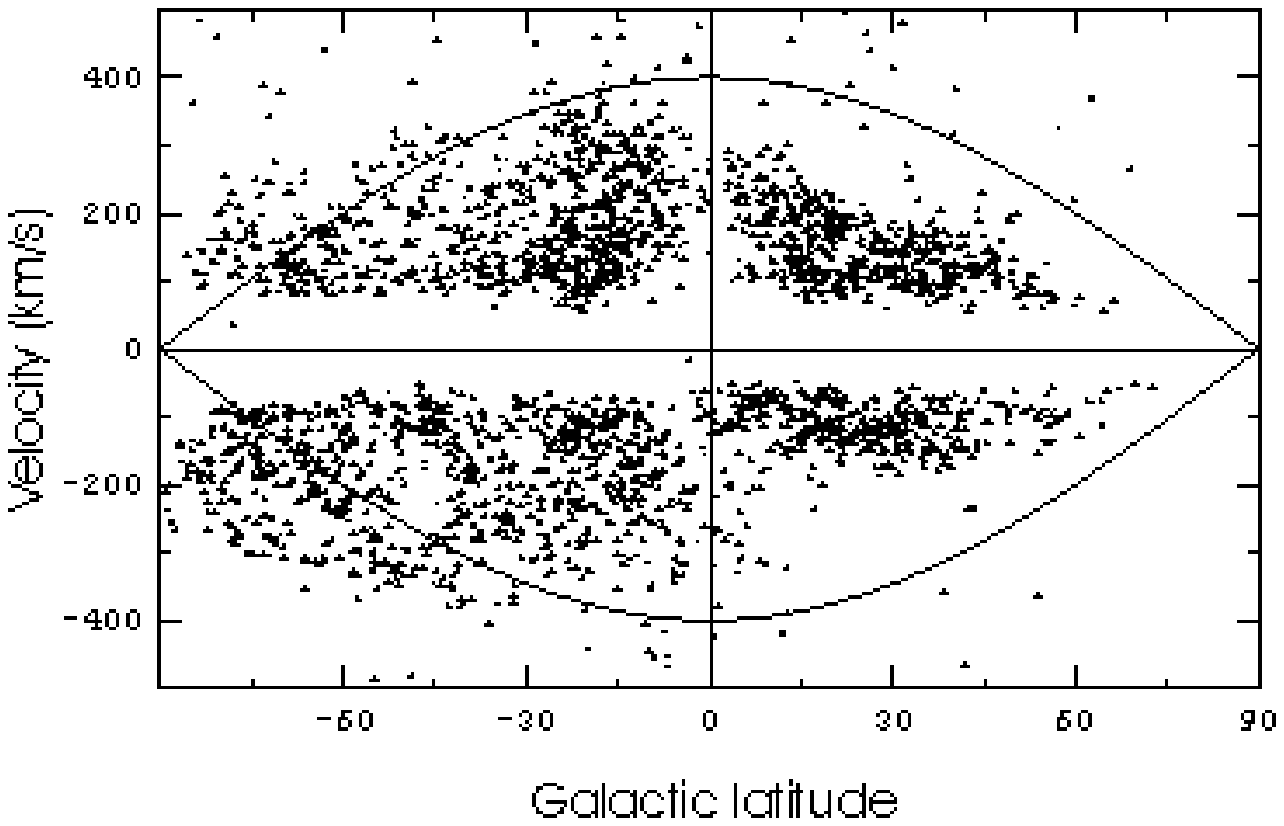}
      \caption{Same as Fig. ~\ref{Lat-Vel1}, but from  the collection of data used in Fig. ~\ref{DataSky2}.
              }
   \label{Lat-Vel2}
   \end{figure}
The enclosing curve  of the longitude-velocity distribution (Figs. ~\ref{Lon-Vel1} and  ~\ref{Lon-Vel2}) is 
approximately sinusoidal and that of the latitude-velocity distribution (Figs. ~\ref{Lat-Vel1} and ~\ref{Lat-Vel2}) 
follows a cosine law. In consequence, the radial velocity distribution of the HVCs can be 
represented  by the following formula:
\begin{equation}
\rho=-V \, sin(\ell) \, cos(b),
\label{velocitylaw}
\end{equation}
where $V \simeq  450\, \mathrm{km\, s^{-1}}$.
This distribution law admits a simple  interpretation (or ``inversion''), namely: it  reflects a 
roughly uniform
  flow of HVCs that  comes approximately from galactic co-ordinates $\ell_{0}=90\degr$ and $b_{0}=0\degr$
 and that encloses the whole Galaxy. If $V_{\sun}$ and  $V_{f}$ are the velocities of the
 Sun and of the HVC
 flow with respect to the Galactic center respectively, the radial velocities of the HVCs with respect to the LSR  are given
 by   
\begin{equation}
\rho=-V_{\sun}\, sin(\ell)\, cos(b)-V_{f}\, cos(\ell-\ell_{0})\, cos(b-b_{0}).
\label{velocitylaw2}
\end{equation}
Substituting the values of $\ell_{0}$ and $b_{0}$ into  Eq. (\ref{velocitylaw2}), we obtain 
 Eq. (\ref{velocitylaw}) with $V=V_{\sun}-V_{f}$. The velocity of the HVC flow is of the order of
 the Solar velocity
 ($220\, \mathrm{km\, s^{-1}}$), and has the same direction, but its sense is reversed. In other words,
 the Sun is immersed in the  flow of HVCs,
 moving counter stream. This flow of clouds seems to circulate largely in the halo or in the outskirts 
of the Galaxy. In the next section, we will find a clue that makes it possible  to connect  the  flow of HVCs with 
the Magellanic Clouds and the Magellanic Stream.

\section{Model for the origin and dynamical  evolution of the high velocity clouds}

In this section we will try to determinate the probable origin of the HVC flow described in Sect. 2  and 
the initial conditions of the phenomenon; with this we will construct a model to study the trajectories of
 the  HVCs
throughout the halo and the disk of the Galaxy. Our first objective is to show that the Magellanic Clouds have 
 played a central role
in the origin of the HVCs. For this purpose, we need to calculate the orbital plane of the Magellanic Clouds.
 We adopt throughout this paper a Cartesian coordinate system $(X,Y,Z)$ with the origin at the Galactic center,
 the X-axis pointing in the direction of the Sun's Galactic rotation, the Y-axis pointing in the direction 
from the Galactic center to the Sun, 
and the Z-axis pointing toward the Galactic north pole. 
 Since the Magellanic Clouds are
 subject to a central force due to the Galactic halo (we will ignore the forces of the galactic disk),
 their orbital plane
can be characterized   by the vector normal to  this  plane, namely  $\vec{N} =  \vec{r}\times\vec{\dot{r}}$, where 
 $\vec{r}=(X,Y,Z)$ is  the position and $\vec{\dot{r}}= (\dot{X},\dot{Y},\dot{Z})$  
 is the  velocity vector of 
the center of  mass of the combined Clouds  at the present epoch
 in the Galactocentric rest frame. We use the  observational parameters of the  LMC,  which are better 
determined,  as representative of  the Magellanic Clouds as a whole. We adopt the
 following   parameters for  the LMC: $\ell=280\degr.5$ and $b=-32\degr.9$  for the Galactic coordinates,
 $50.1\,\mbox{kpc}$  for the distance from the Sun,  $262.2\,\mathrm{km\, s^{-1}}$  for
 the heliocentric systematic velocity (van der Marel et al. \cite{marel}), $\mu_{\alpha}=0.00486 \arcsec\, \mathrm{yr^{-1}}$ and 
$ \mu_{\delta}=0.00034 \arcsec\,\mathrm{yr^{-1}}$ for 
the mean proper motion (see Table 1 of van der Marel et al. \cite{marel}).
 To correct for the reflex motion  of the Sun and to obtain the positions and velocities  in 
the Galactocentric frame of rest, we use the IAU values $R_{\sun}=8.5\,\mbox{kpc}$ and $V_{\sun}=220\,\mathrm{km\, s^{-1}}$
 and the ``basic'' solar motion of $16.5\,\mathrm{km\, s^{-1}}$  towards the direction of  $\ell=53\degr$ and 
 $b=25\degr$ (Allen \cite{allen}).
 With these adopted values,  we calculate   the position and velocity vectors of the LMC  
\begin{equation}
\vec{r}_\mathrm{LMC}=(-41.30, 1.29, -27.43)\, \mbox{kpc}
\label{LMCparameters1}
\end{equation}
and
\begin{equation}
\vec{\dot{r}}_\mathrm{LMC}= (-218, 66, 187)\,\mathrm{km\, s^{-1}}.
\label{LMCparameters2}
\end{equation}
The resulting plane of the orbit of the LMC is represented in Fig. ~\ref{plano-orbita}. 
 The angle between the orbital plane and the plane perpendicular to the Galactic disk is only of the order of $10\degr$. 
i.e  the plane of the LMC orbit  is almost perpendicular to the Galactic plane. The Galactic longitude 
of the  nodal line is nearly  $82\degr$, and thus the orbital plane is almost coincident with the $X$-$Z$ plane 
 (cf. Fig.1 of  Murai \& Fujimoto \cite{murai}).  Therefore,  the three-dimensional velocity vectors 
representing the HVC flow described in Sect. 2 are nearly parallel to the plane 
of the orbit of the Magellanic Clouds. This indicates a dynamical  relationship  between
the HVCs and the Magellanic Clouds. Our working hypothesis will be  that the HVCs were ejected from the Magellanic Clouds
as a result of internal processes in the Magellanic Clouds, activated likely by the collision of the LMC and SMC.  
\renewcommand{\thefigure}{\arabic{figure}}
\begin{figure}
 \includegraphics[width=\textwidth]{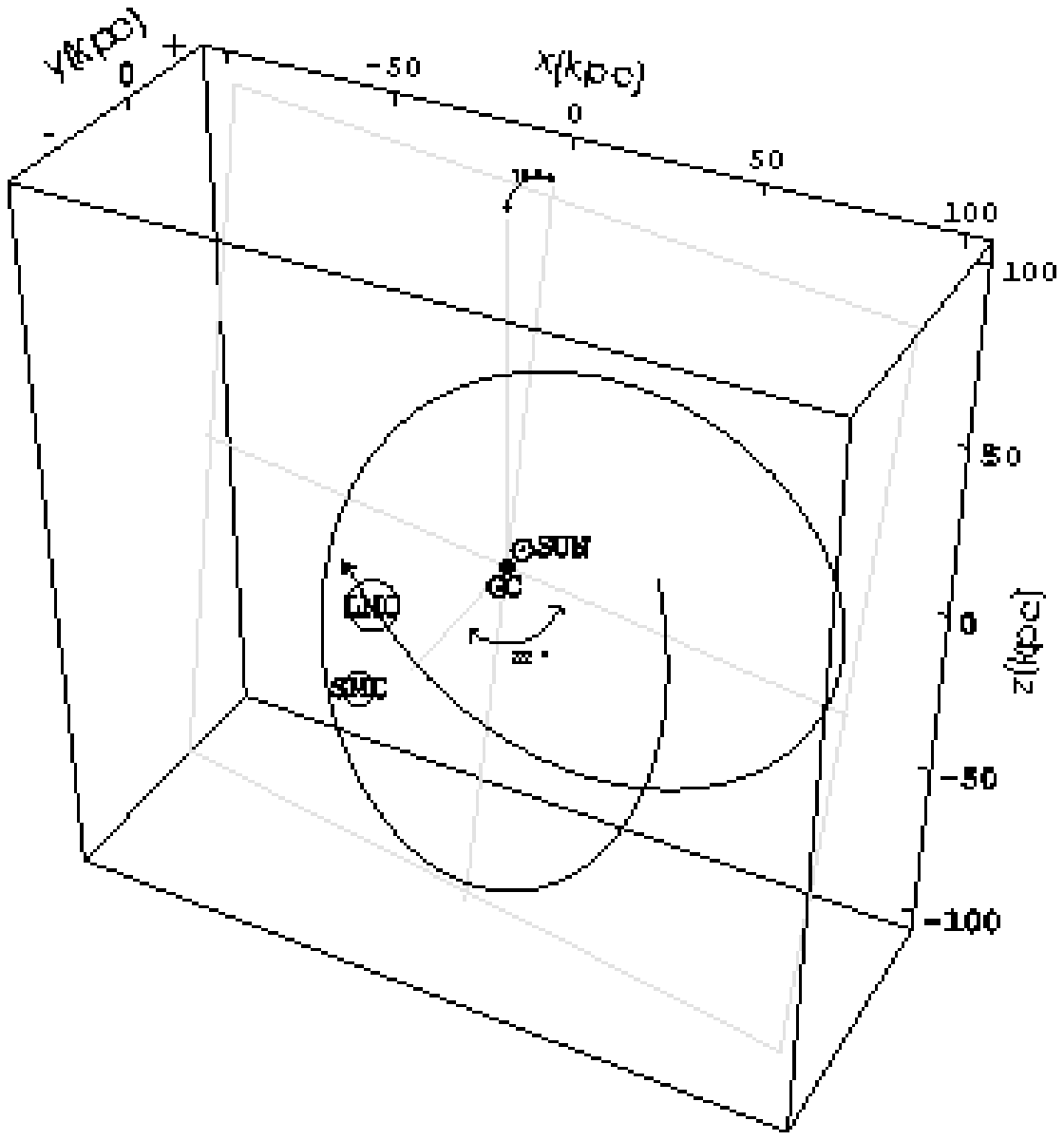}
      \caption{Locations of the LMC, the SMC and the Sun in Galactocentric coordinates $(X,Y,Z)$. The past orbit of the LMC 
is represented  by the solid curve. The position angles of the orbital plane are indicated.  
              }
         \label{plano-orbita}
   \end{figure}

\subsection{Basic equations of the model}
The basic problem we should consider here  is the motion of a test particle representing an HVC in
the gravitational fields  of the Galaxy and the Magellanic Clouds, and friction forces due 
to the gaseous disk of the Galaxy. 

In our computations, we use the spherical gravitational potential of the halo 
$\psi=-V_{c}^{2} \ln r$ for
the Galaxy (Murai \& Fujimoto \cite{murai}; Lin \& Lynden-Bell \cite{lin}), which gives a
 flat rotation curve with a constant circular velocity, $V_{c}=220\,\mathrm{km\, s^{-1}}$,  so  
that the gravitational force of the Galaxy exerted on a particle of unit mass
 is $-V_{c}^{2} \frac{\vec{r}}{r^{2}}$. For the Magellanic Clouds, we use the gravitational potential of
  the LMC 
and ignore that of the SMC. We consider the SMC as another test particle (see Sect.3.2).  We assume that 
the LMC has a Plummer-type potential with an effective radius $K= 3\,\mathrm{kpc}$ (Murai \& Fujimoto \cite{murai}),
giving a gravitational force 
per unit mass of $ -\frac{G M_\mathrm{LMC}  \vec{r'}}{(r'^{2}+K^{2})^{3/2}}$, where   $\vec{r'}(= \vec{r}-\vec{r}_\mathrm{LMC})$ denotes 
the position of the test particle with respect to the LMC, and  $M_\mathrm{LMC}$ is the total mass of the LMC  estimated at   $0.87\times 10^{10}\,\mathrm{M}_{\sun}$ (van der Marel et al. \cite{marel}).

 The friction force  per unit mass that acts  on an HVC can be expressed as
 $\vec{F}_\mathrm{d}=-  \varepsilon\frac{n S}{m} v_\mathrm{r}\vec{v}_\mathrm{r}$, where $\vec{v}_\mathrm{r}$ is the relative velocity between  
the HVC and the gaseous medium  in which the cloud moves, $n$ is the density of the gaseous medium at the position of the HVC,  
S is  the head cross-section of the HVC,
 m is the HVC mass, and $\varepsilon $ is   a dimensionless quantity  $\simeq 1$. A measure of the ratio  $\frac{m}{S}$ is 
the column density $N_\mathrm{H}$ of the HVC being studied, an observable quantity. We adopt a mean $N_\mathrm{H}$ of $7.2\times 10^{18}\,
\mathrm{atoms\, cm^{-2}}$ for the HVCs.   In our simplified model, the Galactic  halo is empty of gas, thus 
 $n$ depends only upon
the density distribution of the gaseous disk of the Galaxy. The density $n$ may  be conveniently
 expressed in cylindrical coordinates as   
$n(R,z)=\frac{\sigma(R)}{2 h(R)} \exp[- \frac{\mid z \mid}{h(R)}] $, where $\sigma(R)$ and $h(R)$ are   
the surface density of interstellar HI   and the scale height of the  thickness of the HI gas layer at $R\, 
(=\sqrt{x^{2}+y^{2}})$, respectively. We denote  by $x,y$ and $z$   the Cartesian components of 
the position vector $\vec{r}$ of the test particle or HVC. From a fit of  Wouterloot et al.'s
 (\cite{wouterloot}) Table 1 we derive $\sigma(R)=1.0875\times 10^{21} \exp [-(\frac{8.24-R}{1.06\times8.24})^{2}]\,\mathrm{atoms\, cm^{-2}}$
 and $ h(R)=105.8 \exp [(\frac{3.37-R}{4.23\times3.37})^{2}]\,\mathrm{pc}$ with $R$ in kpc. Assuming that the gas of
 the Galactic disk rotates
 circularly with  constant  velocity  $V_{c}$, the velocity vector of the gaseous medium at the position of the HVC is
$(\frac{V_{c}}{R}y, -\frac{V_{c}}{R}x, 0 )$ and hence  $\vec{v}_\mathrm{r}= (\dot{x}-\frac{V_{c}}{R}y, \dot{y}+\frac{V_{c}}{R}x,
 \dot{z})$. Under these conditions, the equations of motion of a test particle  are 
\begin{equation}
\ddot{\vec{r}} + \frac{V_{c}^{2} \vec{r}}{\vec{r}^{2}}  + 
\frac{G M_\mathrm{LMC} (\vec{r}-\vec{r}_\mathrm{LMC})}{ (\mid \vec{r}-\vec{r}_\mathrm{LMC} \mid^{2} + K^{2})^{3/2}} - 
 \vec{F}_\mathrm{d}(\vec{r},\dot{\vec{r}})= 0.  
\label{eqmovimiento1}
\end{equation}
To solve this system of equations, we need first the solution of the equations of motion of the LMC 
in the gravitational field of the Galaxy, namely:
\begin{equation}
\ddot{\vec{r}}_\mathrm{LMC} + \frac{V_{c}^{2} \vec{r}_\mathrm{LMC}}{\vec{r}^{2}_\mathrm{LMC}}= 0.  
\label{eqmovimiento2}
\end{equation}

\subsection{The collision between the Large and Small Magellanic Clouds}
Several authors agree  that the Magellanic clouds had a  close encounter 
(separation less than 3 kpc) 
between 200 and 500 Myr  ago (Murai \& Fujimoto \cite{murai}; Gardiner et al. \cite{gardiner2}; Heller \& Rohlfs \cite{heller};
 Moore \& Davis \cite{moore};
 Gardiner \& Noguchi \cite{gardiner}). The morphological peculiarities
of the Clouds such as  the widely scattered distribution of HII regions around the optical bar of the LMC and
the severe fragmentation of the SMC could be understood in terms of this collision.

The encounter time $t_\mathrm{e}$ of the Clouds is 
an important parameter in our model, and we want to calculate  its value congruently 
with the other parameters  adopted above. The orbit of the LMC can be computed  straightforwardly  by means of 
 the numerical integration of Eq.~(\ref{eqmovimiento2}) with the initial conditions given by 
 Eqs.~(\ref{LMCparameters1}) and (\ref{LMCparameters2}) (see Fig. ~\ref{plano-orbita}). For simplicity we assume  that the orbit of the LMC
 is not essentially altered by the presence of the SMC because of its lesser mass.  Knowing  the present position and velocity of
 the center of mass of the SMC, we  compute 
 the orbit of the SMC  by using Eq.~(\ref{eqmovimiento1}) with  $\vec{F}_\mathrm{d}=0$  and the solution of  the orbit of the LMC.
 Thus we  obtain the  time of the closest approach of both orbits. However, the position and velocity  of the SMC 
are not known with great precision. Therefore, in the equations we only introduce  the radial component of the SMC velocity,
  based on  the  heliocentric radial velocity  
of $148\,\mathrm{km\, s^{-1}}$ for the SMC that was  accurately measured
with optical and radio-astronomical methods (Hardy et al. \cite{hardy}), and leave as
 unknowns the proper motions and the distance,  
which were used for determining the tangential components of the velocity.      
Solving Eq.~(\ref{eqmovimiento1}) with the help of Eq.~(\ref{eqmovimiento2}), we obtain $\vec{r}_\mathrm{SMC}$ as a function of 
$\mu_{\alpha}$, $\mu_{\delta}$, the distance $d$ and  time $t$. By requiring that
  $\vec{r}_\mathrm{SMC}( \mu_{\alpha}, \mu_{\delta}, d,t)= \vec{r}_\mathrm{LMC}(t)$, we have three equations with three unknowns 
for each chosen 
past time;  their solutions give us $\mu_{\alpha}(t)$, $\mu_{\delta}(t)$ and $d(t)$. Choosing  the set  of theoretical parameters
 compatible with  the  observational ones, we found that $\mu_{\alpha}$, $\mu_{\delta}$  and $d$  should be close to
 $0.00432 \arcsec\,\mathrm{yr^{-1}}$, $-0.00108 \arcsec\,\mathrm{yr^{-1}}$ (cf. Kroupa \& Bastian \cite{kroupa}) and 60.3 kpc (cf. Westerlund \cite{westerlund}) 
 respectively,   and that the collision 
between the Clouds  occurred  570 $\pm 350$ Myr ago. The uncertainty of $\pm 350$ Myr in the value of $t_\mathrm{e}$ was  derived  from  that of  the present velocity  of the LMC (Eq. 4), $\Delta \vec{\dot{r}}_\mathrm{LMC} \approx (\pm 23, \pm 36, \pm 35) \,\mathrm{km\, s^{-1}}$ (see Eq. 55 of van der Marel et al. \cite{marel}), with the assumption  that the rest of the adopted parameters like   the distances to  the  Large and Small Magellanic Clouds  and the gravitational potential used for the halo of 
the Galaxy  are almost exact.  The orbits of the Clouds are displayed in Figs. 
 ~\ref{colision1} and  ~\ref{colision2}.

\setcounter{newctr}{1}
\renewcommand{\thenewctr}{\alph{newctr}}
\renewcommand{\thefigure}{\arabic{figure}-\thenewctr}
 \begin{figure}
   \centering
 \includegraphics[width=\textwidth]{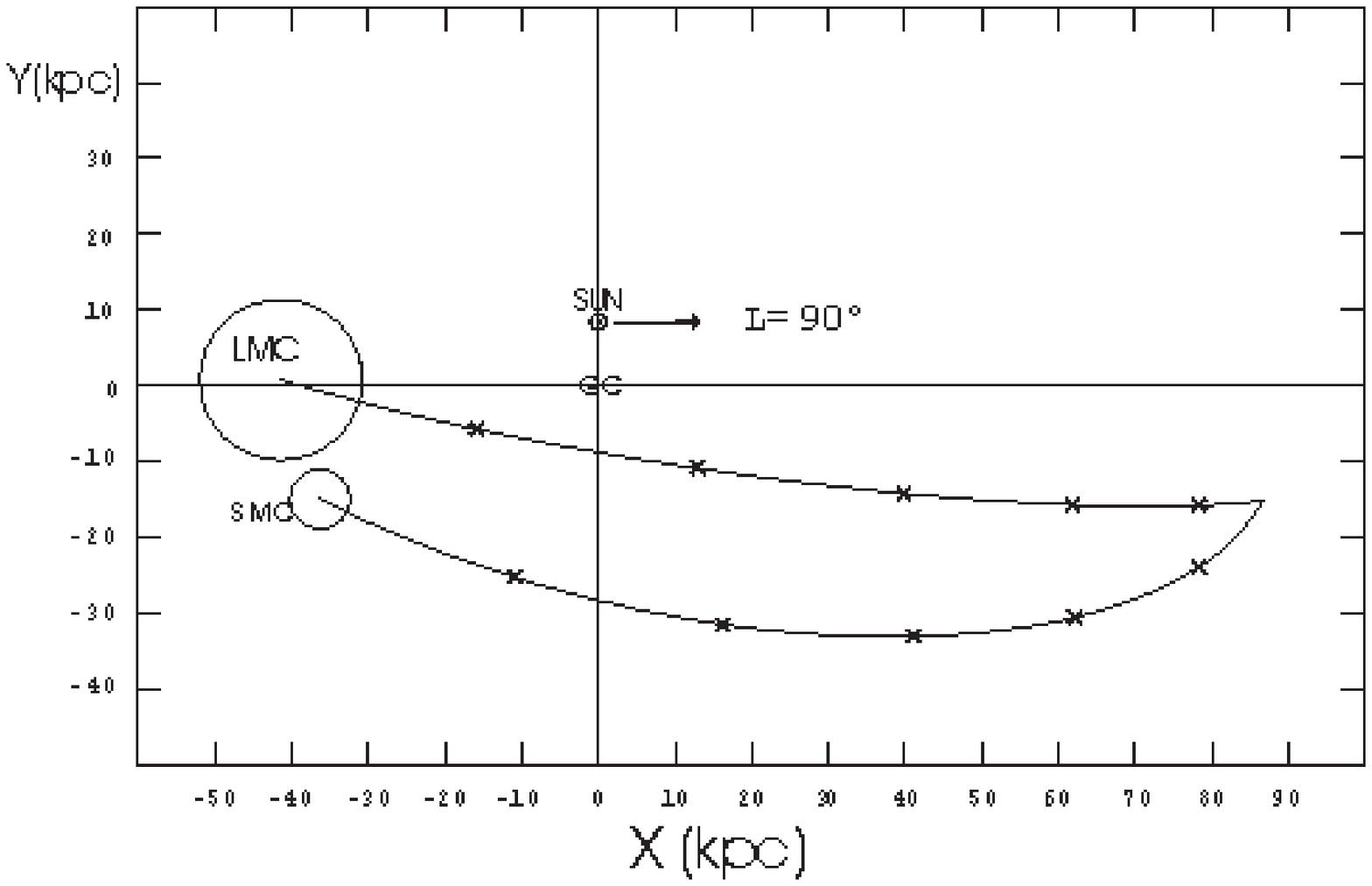}
      \caption{Projection of the past orbits of the LMC and the SMC on the Galactocentric $X$-$Y$ plane.  The LMC and SMC positions  are  marked by crosses  each 100 Myr.
              }
         \label{colision1}
   \end{figure}
\addtocounter{figure}{-1}
\addtocounter{newctr}{1}
\begin{figure}
   \centering
 \includegraphics[width=\textwidth]{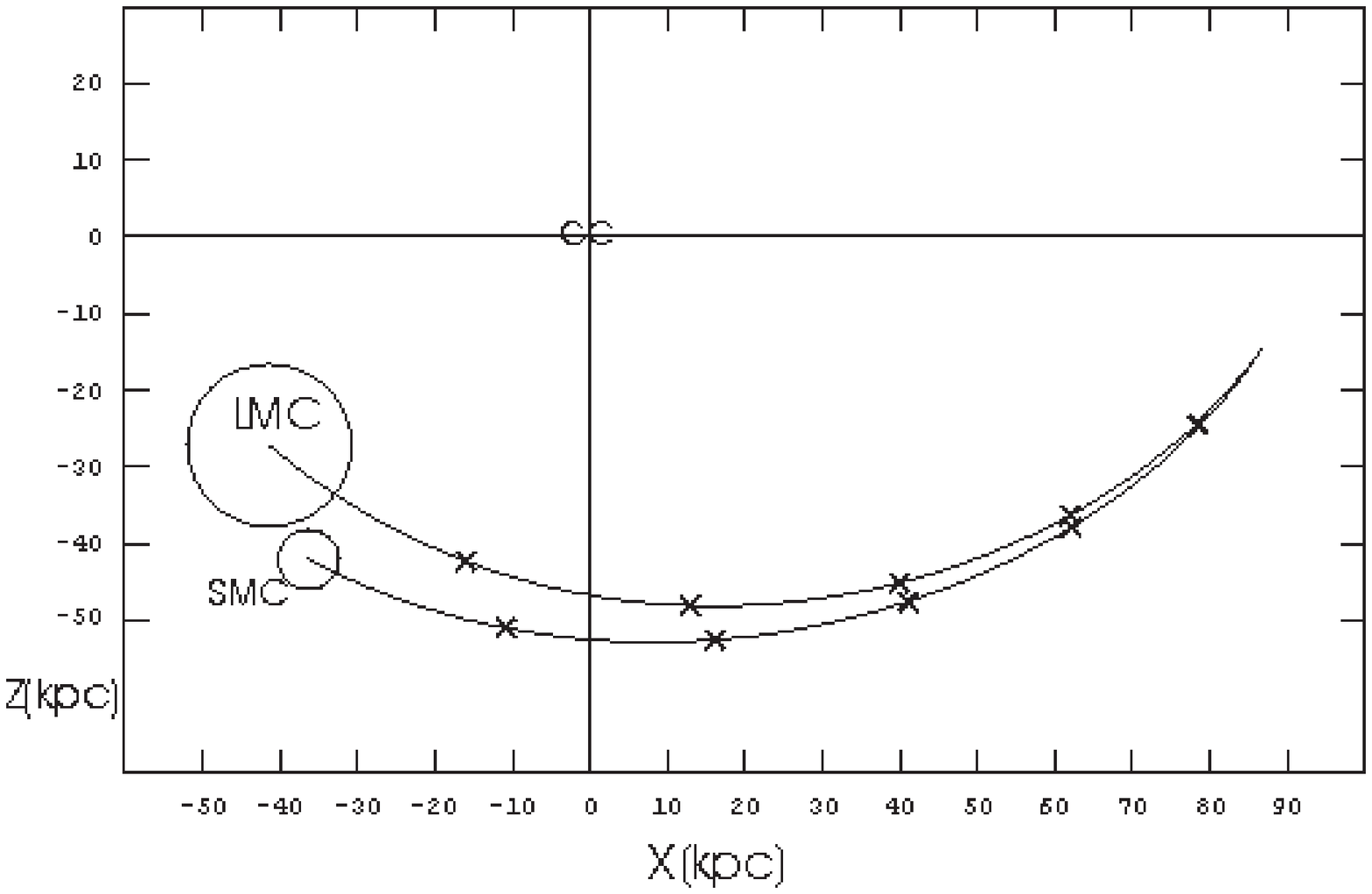}
      \caption{Projection of the past orbits of the LMC and the SMC on the Galactocentric $X$-$Z$ plane.  The LMC and SMC positions  are  marked by crosses  each 100 Myr.
              }
         \label{colision2}
   \end{figure}

The velocity of the LMC at the time of the collision $t_\mathrm{e}=-570\,\mbox{Myr}$ was 
\begin{equation}
\vec{\dot{r}}_\mathrm{LMC} (t_\mathrm{e})= (-100, -12, -141)\,\mathrm{km\, s^{-1}},
\label{Vcolision}
\end{equation}
while that of the SMC was
$\vec{\dot{r}}_\mathrm{SMC} (t_\mathrm{e})= (-100,-167, -155)\,\mathrm{km\, s^{-1}}$. Thus, during the encounter, 
the  velocity of the SMC relative to the LMC  was of order $156\,\mathrm{km\, s^{-1}}$;  i.e. 
very close  to the escape velocity from the LMC ($v_\mathrm{e}=\sqrt{\frac{2GM_\mathrm{LMC}}{K}}\simeq 160\,\mathrm{km\, s^{-1}}$). 
To explain this, we propose that the Magellanic Clouds formed a binary system in the past and that it was broken by 
a sudden mass loss of the system taking place in the epoch of the collision. The collision compressed 
the interstellar gas of the Clouds, inducing a burst of star formation. The massive stars of this new stellar generation expelled  the surrounding  interstellar matter in a few millions of years at  velocities 
greater than the escape velocity of the system,
 through supernova explosions and 
strong stellar winds. A part of the material ejected from the Clouds at that time formed the HVCs. 
The HVCs left likely the Magellanic Clouds as magnetized clouds of plasma (eg. Olano \cite{olano1}). Magnetic fields in HVCs have been detected  by measuring  Zeeman splitting of the 21-cm HI line (Kaz\`{e}s et. al. \cite{kazes}). 
This magnetic self-confinement of the HVCs 
could  explain  the HVC longevity as discrete entities. This idea is coherent with the fact that the HVCs are highly ionized (Sembach et al. \cite{sembach}; Lehner et al. \cite{lehner}; Sembach et al. \cite{sembach2}). The HVCs are still generally thought of as neutral entities. However, at an early stage they were probably fully ionized. Subsequently,  the evolution of their physical conditions permitted  the recombination of part of the protons and electrons.  On the other hand, the explanations generally given for the cloud confinement are that they are either bound by gravitationally dominant dark matter or  confined by an external pressure provided by   a hot gaseous Galactic halo (Espresate et al. \cite{espresate}; Stanimirovi\'{c} et al. \cite{stanimirovic}; Sternberg et al. \cite{sternberg}).

\subsection{Initial conditions of the high velocity clouds}
We conjecture that the HVCs were ejected from the Magellanic Clouds as the consequence of very 
energetic interactions of  a large number of stars born during the collision of the Clouds  
and the cloudy interstellar medium  of the Clouds. The relevant initial conditions, namely   the positions, 
velocities and  times  of  the ejections of the HVCs, should    therefore be characterized 
by  distribution functions.
For simplicity we assume  that the HVCs were ejected from a relatively small region of the Clouds,
at the central point of the collision, during a short period of time, i.e. at  $t_\mathrm{e}=$-570 Myr.
The position vector of the central point of the collision (see Sect. 3.2 and Figs. 
 ~\ref{colision1} and ~\ref{colision2}) is then 
\begin{equation}
\vec{r}_\mathrm{LMC}(t_\mathrm{e})=(86.39,-15.53, -15.68)\,\mathrm{kpc},
\label{Pcolision}
\end{equation}
and the initial position vector for all HVCs  is $\vec{r}_{0}  =\vec{r}_\mathrm{LMC}(t_\mathrm{e})$.
For the initial systematic or mean velocity of the HVCs we adopt 
the velocity  of the mass center of the LMC at the time $t_\mathrm{e}$, which is  given by  Eq.~(\ref{Vcolision}).  
 The errors in  Eqs.~(\ref{Vcolision}) and (8), i.e. in the position of the LMC in phase space at  $t_\mathrm{e}$,  are $ \Delta \vec{\dot{r}}_\mathrm{LMC} (t_\mathrm{e}) \approx (\pm 3.87, \pm 0.72, \pm 1.03 )\times  \mid  \Delta \vec{\dot{r}}_\mathrm{LMC}\mid \, \mathrm{km\, s^{-1}}$  and $\Delta \vec{r}_\mathrm{LMC}(t_\mathrm{e}) \approx (\pm 0.79, \pm 0.20, \pm 1.21)\times  \mid  \Delta \vec{\dot{r}}_\mathrm{LMC}\mid \, \mathrm{kpc}$, obtained by propagation of the errors in $\vec{\dot{r}}_\mathrm{LMC}$ and $t_\mathrm{e}$ (Sect. 3.2). 

Thus the initial total velocity of an HVC can be written as 
\begin{equation}
\vec{\dot{r}}_{0}=\vec{\dot{r}}_\mathrm{LMC} (t_\mathrm{e})+ \vec{v}_\mathrm{p},
\end{equation}
 where  $\vec{v}_\mathrm{p}$ is 
the initial peculiar velocity of the HVC. The number of HVCs in the initial velocity volume  $d\vec{v}_\mathrm{p}$ at  $\vec{v}_\mathrm{p}$ is
$dN= f(\vec{v}_\mathrm{p}) H(v_\mathrm{p}-v_\mathrm{e}) d\vec{v}_\mathrm{p}$ at time $t_\mathrm{e}$. We assume  that the distribution function $f(\vec{v}_\mathrm{p})$ follows 
 Schwarzschild's ellipsoidal law:

\begin{equation}
f(\vec{v}_\mathrm{p})=\frac{N}{(2\pi)^{3/2} \Sigma_{1}  \Sigma_{2}  \Sigma_{3}} 
\exp{-\frac{1}{2}( (\frac{\dot{x}_\mathrm{p}}{\Sigma_{1} })^{2} +  (\frac{\dot{y}_\mathrm{p}}{\Sigma_{2} })^{2} +
(\frac{\dot{z}_\mathrm{p}}{\Sigma_{3} })^{2} ) )},
\label{Schwarzchild}
\end{equation}
where 
$\Sigma_{1}$, $\Sigma_{2}$ and $\Sigma_{3}$
 are  the three velocity dispersions in the directions of the three co-ordinate axes and
    $\dot{x}_\mathrm{p}$, $\dot{y}_\mathrm{p}$ and $\dot{z}_\mathrm{p}$ are the components of the peculiar velocity vector 
 $\vec{v}_\mathrm{p}$. The Heavyside function $H$, which  is 1  for $v_\mathrm{p}-v_\mathrm{e} > 0$ and 0 otherwise, takes 
into account that only the clouds  with velocities  greater than the escape velocity $v_\mathrm{e}$ became  HVCs.
In general form,  the function of initial distribution of the HVCs may be written as
  $F(\vec{r}_{0},\vec{\dot{r}}_{0},t)= \delta(\vec{r}_{0}-\vec{r}_\mathrm{LMC}(t_\mathrm{e})) \delta(t-t_\mathrm{e}) f(\dot{\vec{r}}_{0}-\vec{\dot{r}}_\mathrm{LMC} (t_\mathrm{e}) ) H(\dot{\vec{r}}_{0}-\vec{\dot{r}}_\mathrm{LMC} (t_\mathrm{e})-v_\mathrm{e})$, where $\delta$ is the 
standard  delta function. The total number of HVCs (or the number of test particles used in the simulation) can be expressed
by $ N_\mathrm{HVC}=\int_{0}^{\infty} F(\vec{r}_{0},\vec{\dot{r}}_{0},t)  d\vec{r}_{0} d\vec{\dot{r}}_{0} dt$,
 which makes it possible to determine
$N$ of Eq.~(\ref{Schwarzchild}).

\section{Results of the  model and comparison with the observations}

To reproduce 
the main features of 
the position and velocity distributions of the HVCs we calculated the orbits of about 850 test particles, which  represent  the HVCs. Distributing these test particles in  phase space 
according to the distribution functions  defined in Sect. 3.3, we can assign an initial position and velocity to each test particle, and calculate its present position and velocity by means of the equations of motion given  in Sect. 3.1. The three dispersions of the velocity ellipsoid in Eq.~(\ref{Schwarzchild}) 
are the free parameters of the model. Their values  can  be estimated  by 
comparing the predictions  of the model with the observations.  We found that the case particular of a three-dimensional 
Gaussian distribution  is 
compatible with the observations, namely  $\Sigma_{1}=\Sigma_{2}=\Sigma_{3}= \Sigma= 120\,\mathrm{km\, s^{-1}}$. Thus,  $\Sigma$ is the only free parameter of the model.

To consider the possibility of non-isotropic expulsion of material, we  tried   other initial ellipsoidal distributions with different  values and spatial orientations of their velocity dispersions with respect to the co-ordinate axes, but none of them could  produce better results than  those of a Gaussian distribution. This simple  distribution   reproduces fairly well  the kinematics and sky distributions  of the Magellanic stream, the ``classic'' stream  and the leading stream (Putman et al. \cite{putman1}), as well as those of the rest of the HVC population. Certainly, quantitative comparisons  between theoretical and observed distributions in the way developed by Saha (\cite{saha}) would be desirable. However, it is beyond  the scope of the present work. A proof of the robustness of  our method is that  the results of the simulations are rather insensitive to the uncertainties of the time of the collision, $\Delta t_\mathrm{e}$, and the velocity, $ \Delta \vec{\dot{r}}_\mathrm{LMC} (t_\mathrm{e})$,  and position, $\Delta \vec{r}_\mathrm{LMC}(t_\mathrm{e})$,  of the centroid of the initial Gaussian distribution.      

 The results of the model are shown in Figs.  ~\ref{sky-teorica}, ~\ref{Lon-Vel-teorica},
~\ref{Lat-Vel-teorica}, ~\ref{espacial} and ~\ref{espacialXZ}.
 Comparing Fig.  ~\ref{sky-teorica} with Figs. ~\ref{DataSky1} and  ~\ref{DataSky2}, 
Fig. ~\ref{Lon-Vel-teorica} with  Figs.~\ref{Lon-Vel1} and ~\ref{Lon-Vel2}, and Fig. ~\ref{Lat-Vel-teorica} with 
Figs.~\ref{Lat-Vel1} and ~\ref{Lat-Vel2}, we see that the theoretical distributions are in good agreement with 
the corresponding observed ones. An exception is a group of  HVCs that lies in a region centered at $\ell=180^{\circ}$, $b=-20^{\circ}$, 
 whose velocities are not explained by the model. 
However, we should remember the simplifications and uncertainties of the observational parameters 
introduced into the model, as well as  the probable existence  of other sources of HVCs, apart from the 
Magellanic Clouds. Studies of HVCs in external galaxies (see Schulman et al. \cite{schulman}; Wakker \& Woerden \cite{wakker-woerden}; Jim\'{e}nez-Vicente \& Battaner \cite{battaner};  Miller et. al. \cite{miller}) will enable us to make a comparison with the HVC system associated with our own Galaxy and  to decide the relevance  of the other possible sources of HVCs, such as  Galactic fountains (Bregman \cite{bregman};  Houck \& Bregman \cite{houck}). The part of the HVC population of Magellanic origin having  LSR  radial velocities between  -100 and +100 $\mathrm{km\, s^{-1}}$ and lying at low Galactic latitudes (see Figs.~\ref{Lon-Vel-teorica} and  ~\ref{Lat-Vel-teorica}) should be indistinguishable from the interstellar clouds of the Galactic disk that are parts of supernova shells or that participate simply in the Galactic rotation. 

There is a clear difference between  the two Galactic hemispheres. In the northern hemisphere the population of HVCs is dominated by a few extended, complex structures (the A, C and M objects). When comparing  the observed HVC distributions with the theoretical ones, we should remember  this. An  HVC sampled at regular angular intervals  is  characterized  by a set of neighboring points in the ($\ell$, $b$, $\rho$) space whose  number increases proportionally with  the solid angle subtended by the cloud in the sky (see Sect. 2). The few northern complexes of HVCs  cover a large area of the sky. Hence their densities  of  points  in the observed distributions relative to the densities of points in the corresponding theoretical distributions should be  greater than the relative densities of points  corresponding to the southern HVCs. The results of our simulation agree with this picture (cf. Figs. ~\ref{DataSky1} and  ~\ref{sky-teorica}, and  Figs  ~\ref{Lat-Vel1} and  ~\ref{Lat-Vel-teorica}). According to  our  model, the  explanation of  these  asymmetries between the northern and southern hemispheres is that  in the northern Galactic hemisphere there are  far fewer  HVCs and they lie  at much smaller distances (see Fig. ~\ref{espacialXZ}). Therefore, they subtend  far larger angles on the sky.          
 The importance of the model is that it provides 
a probable   velocity field  and space distribution of  the HVCs in the three 
dimensions (Figs.~\ref{espacial} and ~\ref{espacialXZ}). This 
information  cannot be obtained  from the observations so far.
\setcounter{newctr}{1}
\renewcommand{\thenewctr}{\alph{newctr}}
\renewcommand{\thefigure}{\arabic{figure}-\thenewctr}
 \begin{figure}
   \centering
 \includegraphics[width=\textwidth]{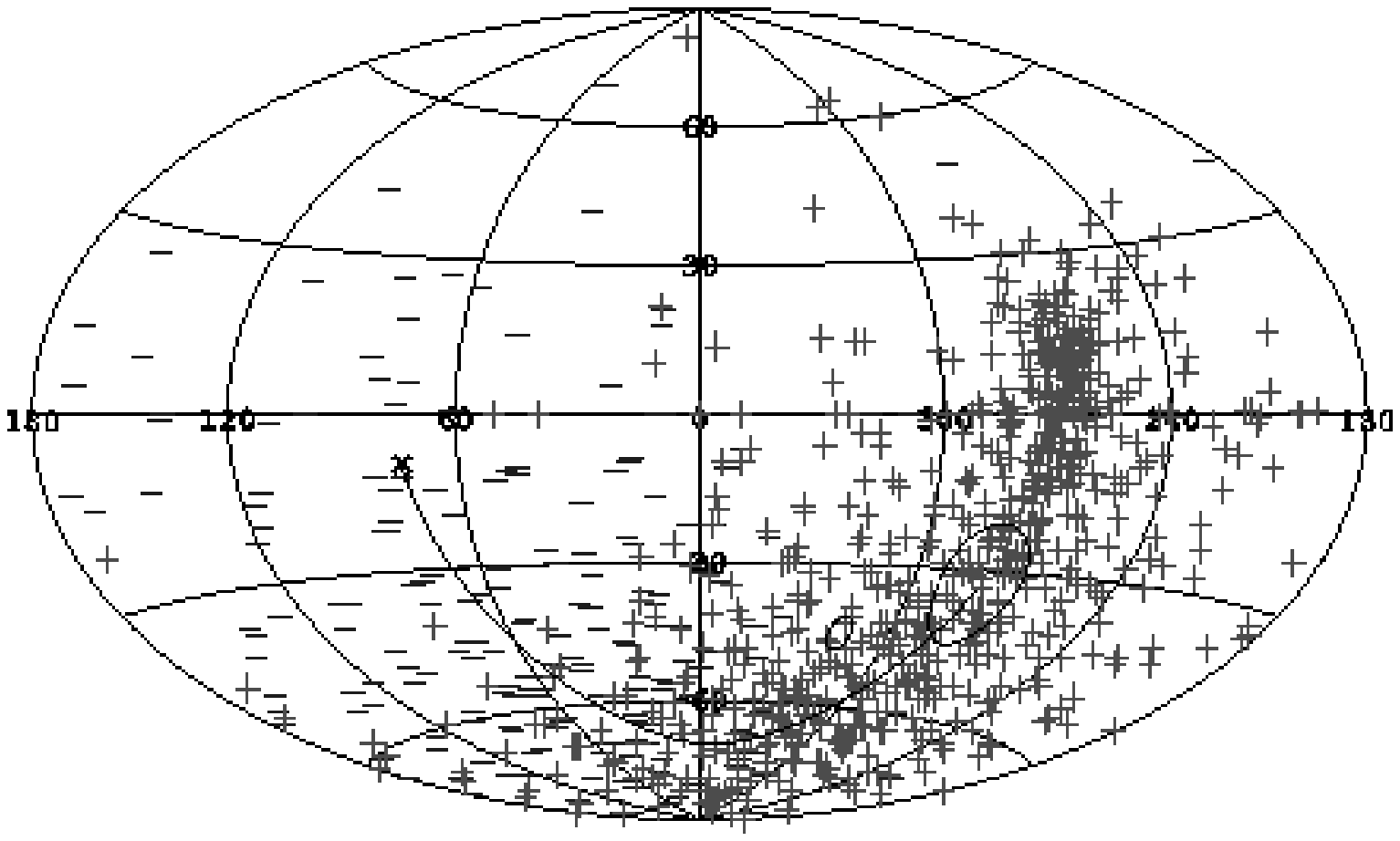}
      \caption{Simulated celestial  distribution of the HVCs: projection of  the positions of the test particles  at present moment  on the sky plane. The plus and minus symbols have the same meanings as in  Fig. ~\ref{DataSky1}. The projections of the past orbit of the LMC (solid curve) and  the present positions of the LMC and SMC (large and small projected circles) are shown. The cross at the end of the projected orbit curve is the projected position of the encounter between the LMC and the SMC. Compare this theoretical distribution with the observed ones (Figs.  ~\ref{DataSky1} and ~\ref{DataSky2}). 
              }
         \label{sky-teorica}
   \end{figure}
\addtocounter{figure}{-1}
\addtocounter{newctr}{1}
\begin{figure}
   \centering
 \includegraphics[width=\textwidth]{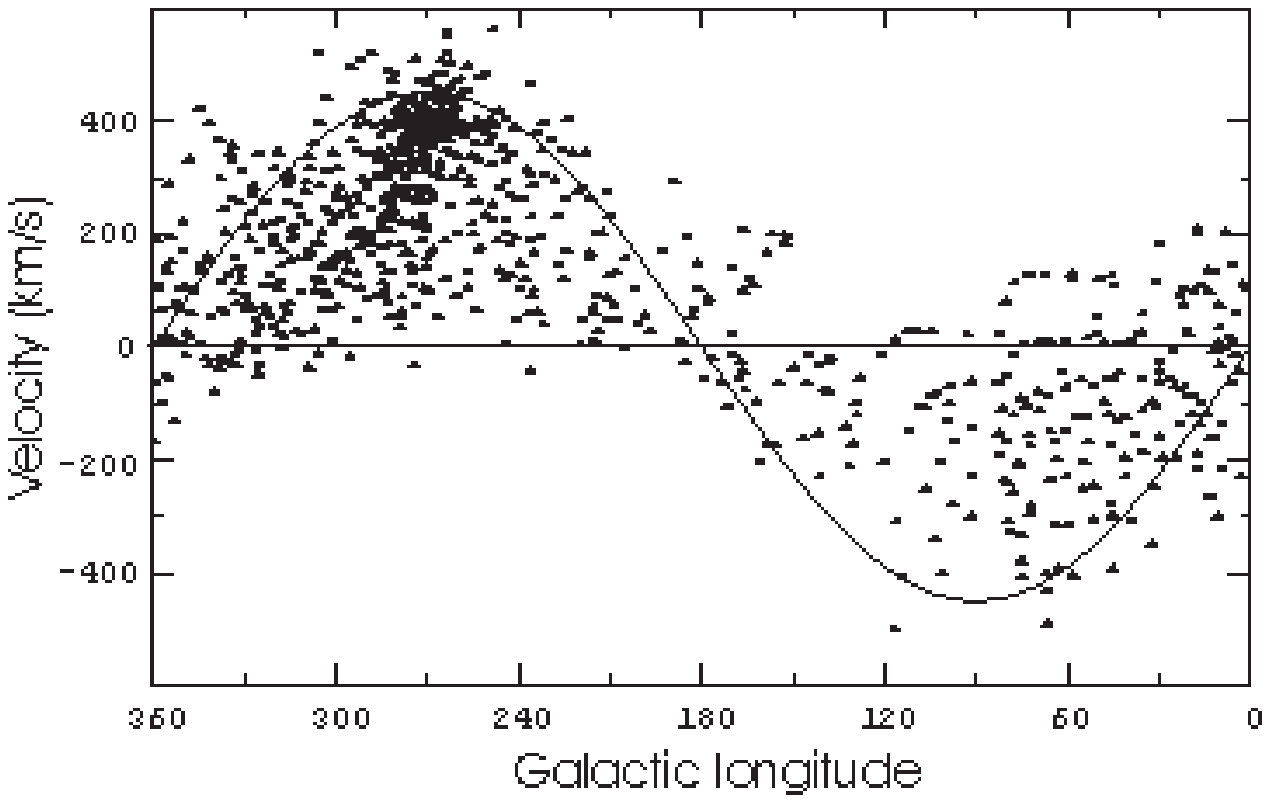}
      \caption{Simulated velocity-longitude relation for the HVCs: radial LSR velocities of the test particles at the present moment as a function of their respective Galactic longitudes. The solid curve has the same meaning as in  Fig.~\ref{Lon-Vel1}. Compare this theoretical relation with the observed ones (Figs. ~\ref{Lon-Vel1} and ~\ref{Lon-Vel2}).
              }
         \label{Lon-Vel-teorica}
   \end{figure}
\addtocounter{figure}{-1}
\addtocounter{newctr}{1}
\begin{figure}
   \centering
 \includegraphics[width=\textwidth]{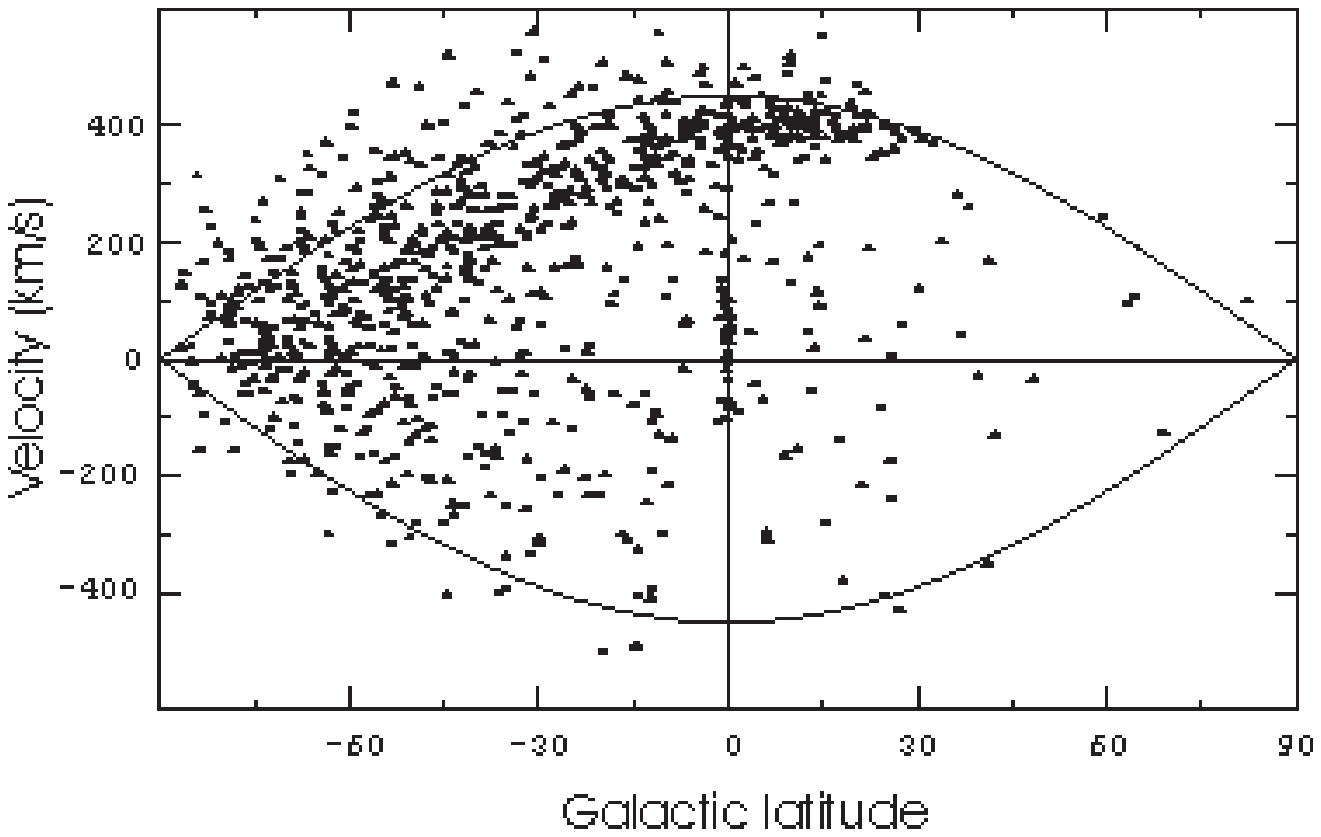}
      \caption{Simulated velocity-latitude relation for the HVCs: present radial LSR velocities of the test particles as a function of their respective Galactic latitudes. The solid curve has the same meaning as in  Fig.~\ref{Lat-Vel1}. Compare this theoretical relation with the observed ones (Figs. ~\ref{Lat-Vel1} and ~\ref{Lat-Vel2}).
              }
         \label{Lat-Vel-teorica}
   \end{figure}
\addtocounter{figure}{-1}
\addtocounter{newctr}{1}
\begin{figure}
   \centering
 \includegraphics[width=\textwidth]{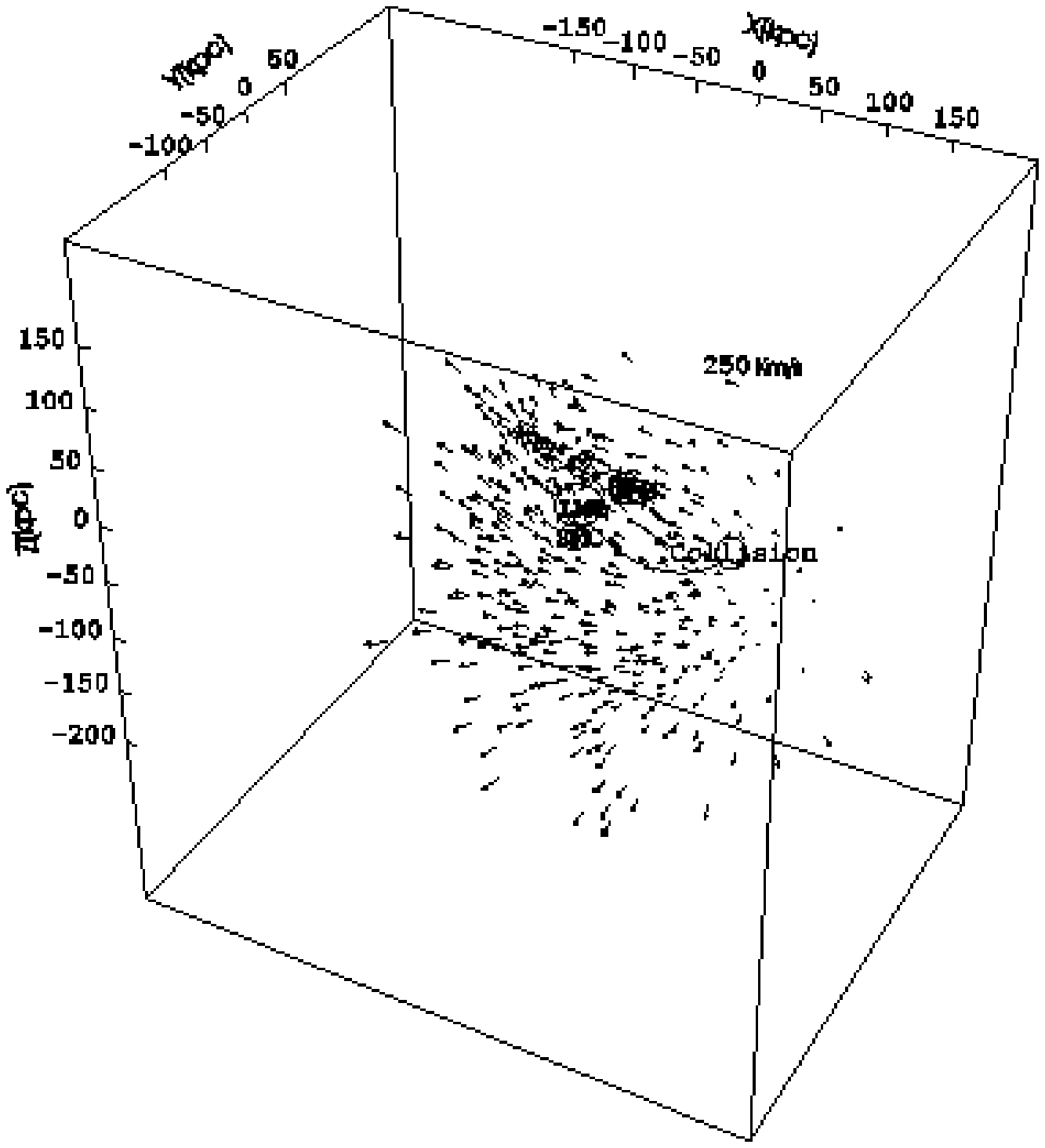}
      \caption{Simulated spatial distribution and velocity field in the Galactocentric rest  frame for the HVCs: Present positions and velocities of the test particles. The test particles with initial velocities  $\mid \vec{v}_\mathrm{p} \mid < 220\,\mathrm{km\,s^{-1}}$ have been omitted in the plot for clarity. The present positions of the LMC and the SMC (circles), the past orbit of the LMC (solid curve) and the point of encounter between the LMC and the SMC (circle) are represented.}
         \label{espacial}
   \end{figure}
\addtocounter{figure}{-1}
\addtocounter{newctr}{1}
\begin{figure}
   \centering
 \includegraphics[width=\textwidth]{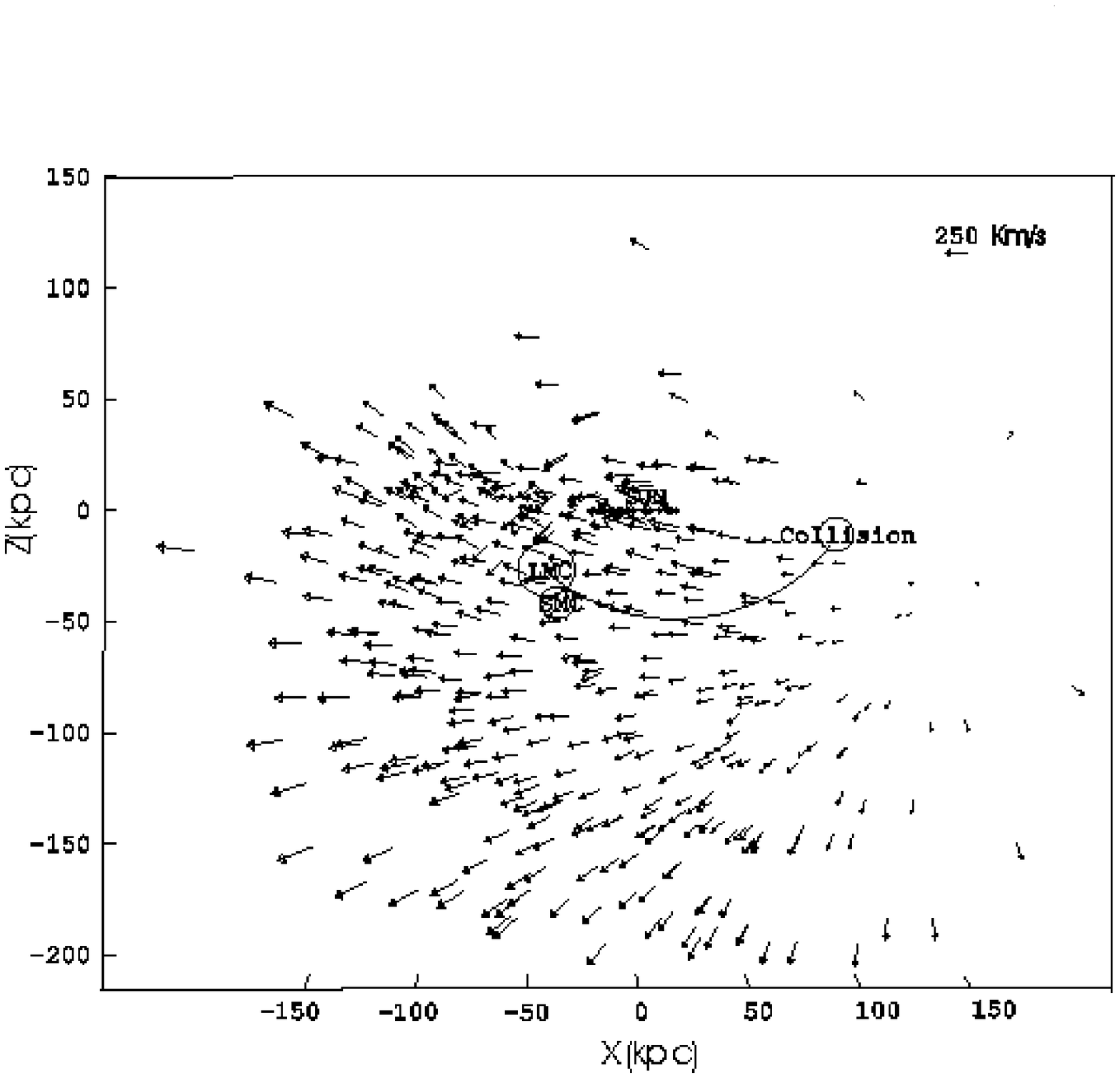}
      \caption{Same as Fig. ~\ref{espacial}, but projected on the Galactocentric $X$-$Z$ plane.
              }
         \label{espacialXZ}
   \end{figure}

The  group of HVCs spreads throughout a  volume of approximately cubic 300 kpc. The mean velocity  of 
the HVC flow is similar to that  of the Magellanic Clouds, since  the HVCs started their trajectories with the
mean or  systematic velocity of the Clouds. To illustrate  
this effect further, we apply the model to a set of test particles  with initial peculiar velocities of the  
same magnitude  $\mid \vec{v}_\mathrm{p} \mid= $cont, but with different directions. At the present time, these test 
particles should form a stream that moves approximately in the same direction and sense as the Magellanic Clouds 
(Fig. ~\ref{cascara1}). In addition,
this cloud of particles is elongated in the direction of the velocity of  rotation around the Galactic center,
 an effect due to  the differential rotation  and well known in the evolution of expanding Galactic shells (e.g.
 Olano \cite{olano2}; Palous, Franco \& Tenorio-Tagle \cite{palous}; Olano \cite{olano3}). Notice that 
the degree of concentration of the particles around the Magellanic orbit  increases as the initial peculiar 
velocity decreases (cf. Figs. ~\ref{cascara1} and   ~\ref{cascara2}). Also, notice that the lower 
the peculiar velocities  of the particles,   the greater the number density of  
particles (or HVCs), according to  the initial distribution function Eq.~(\ref{Schwarzchild}). Both these effects  contribute  
to create in the sky the appearance of the Magellanic Stream (see Fig.  ~\ref{sky-teorica}). 
\setcounter{newctr}{1}
\renewcommand{\thenewctr}{\alph{newctr}}
\renewcommand{\thefigure}{\arabic{figure}-\thenewctr}
 \begin{figure}
   \centering
 \includegraphics[width=\textwidth]{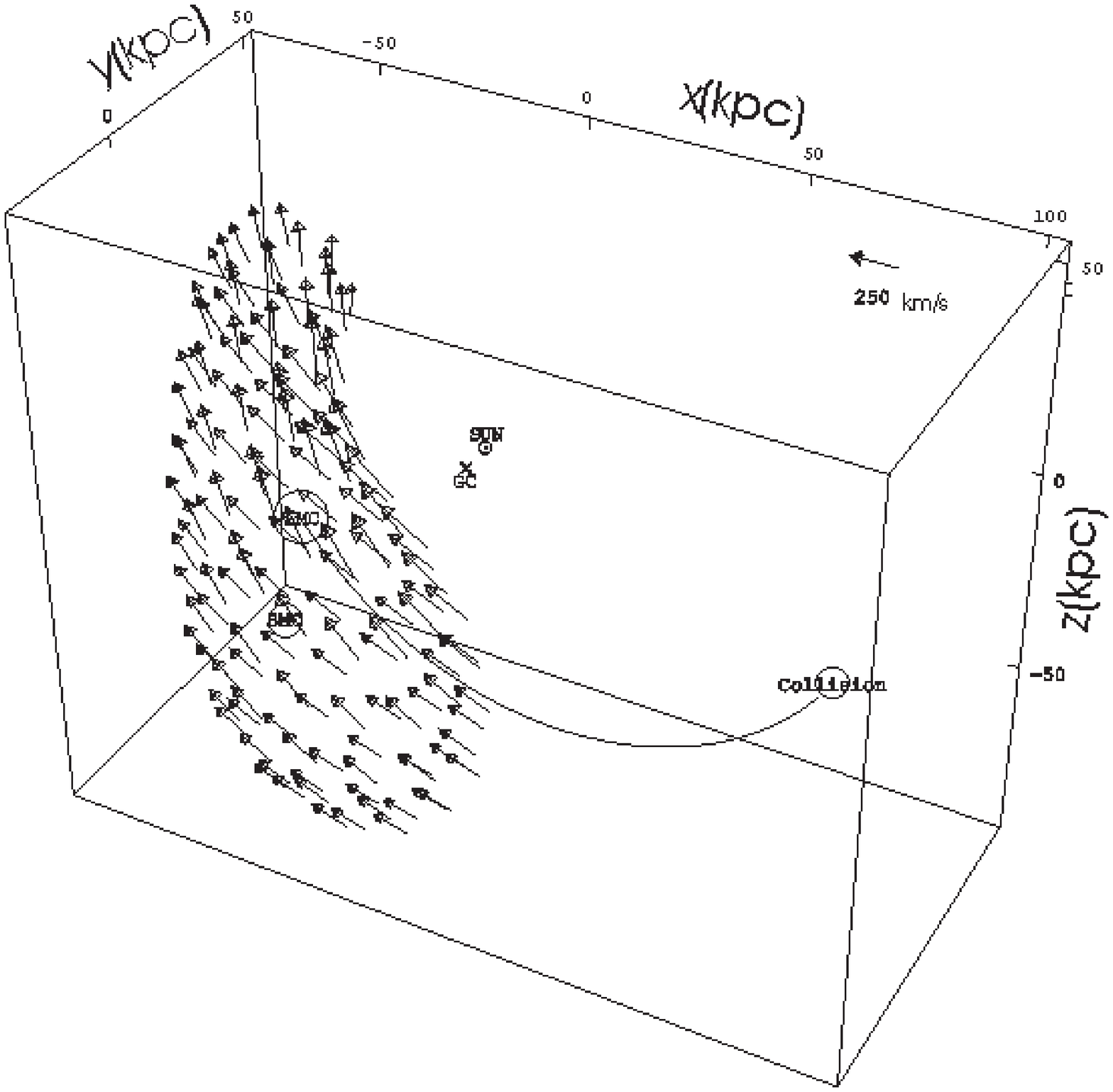}
      \caption{Spatial  distribution and velocity field of the test particles with an initial  isotropic velocity distribution 
characterized by $\mid \vec{v}_\mathrm{p} \mid = 170\,\mathrm{km\,s^{-1}}$.
              }
         \label{cascara1}
   \end{figure}
\addtocounter{figure}{-1}
\addtocounter{newctr}{1}
\begin{figure}
   \centering
 \includegraphics[width=\textwidth]{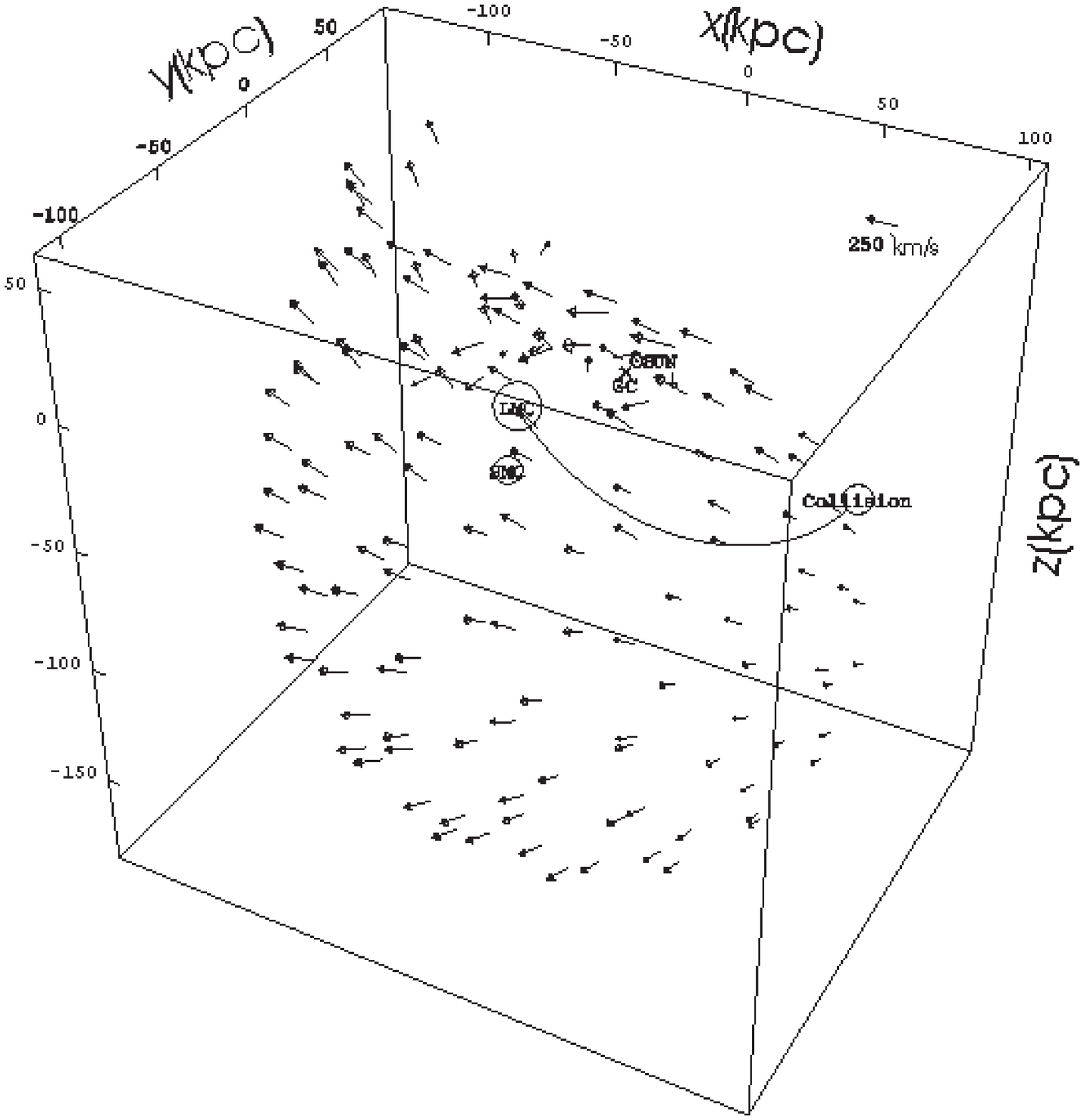}
      \caption{Spatial  distribution and velocity field of the test particles with an initial  isotropic velocity distribution 
characterized by $\mid \vec{v}_\mathrm{p} \mid = 270\,\mathrm{km\,s^{-1}}$.
              }
         \label{cascara2}
   \end{figure}

From the distances to  the HVCs predicted by the model and  the density column  data of Wakker's (\cite{wakker-privado}) list we estimated the total mass of the HVCs by means of the formula 
 $M_\mathrm{t}(\mathrm{M}_{\sun})=6.085 \times 10^{-19}  \overline{d_{i}^{2}} (\sum_{i=1}^{N} N_{\mathrm{H}_{i}})$,   where 
 $\sum_{i=1}^{N} N_{\mathrm{H}_{i}}$  is the sum of the column densities in $\mathrm{atoms\, cm^{-2}}$ of  all  high-velocity profile components
 detected with  a telescope beam of $\sim 0\degr .5$, and  $\overline{d_{i}^{2}}$ is 
 the theoretical  mean squared distance of the HVCs from  the Sun, expressed in $\mathrm{kpc^{2}}$. The results are   
$\sum_{i=1}^{N} N_{\mathrm{H}_{i}}=1.87\times 10^{23}\,\mathrm{ atoms\, cm^{-2}}$ and $\sqrt{\overline{d_{i}^{2}}}=105 \,\mathrm{kpc}$, hence the total HI mass of the HVCs is  
$\sim 1.2 \times 10^{9} \mathrm{M}_{\sun}$. Assuming that the HVCs have as much ionized as neutral hydrogen  (Sembach et al.
\cite{sembach1}), and adopting a factor 1.3  to include  $H_\mathrm{e}$,  we obtain $M_\mathrm{t}= 3.1 \times 10^{9} \mathrm{M}_{\sun}$. Thus the Magellanic Clouds lost about  25 per cent of  their  original mass in the form of
HVCs. This is consistent  with the idea that 
the Clouds constituted a binary system before the last collision, whereas  
the important mass loss  induced by this collision transformed the Clouds into  an unbound system.

 To study  the history  of the Clouds as a binary system we should calculate  the relative motion of 
the SMC and  LMC before the collision. This can be realized 
 by means  of  the equation of  motion resulting from the difference of 
 Eq.~(\ref{eqmovimiento1}) and  Eq.~(\ref{eqmovimiento2}), the boundary conditions at the moment of the collision,
 and a LMC mass  greater than the present one (see Sect. 3.1).
 However, for this  purpose we merely  quote the  pioneering  study on  this topic (Murai \& Fujimoto \cite{murai}). According to it the LMC and the SMC formed a stable binary system during $10^{10}\,\mathrm{yr}$ (see also Yoshizawa \& Noguchi \cite{yoshizawa}). 

Another interesting result of the model is the estimate of the total energy associated with the production
of the HVCs. The total initial kinetic energy of all  HVCs ($\frac{1}{2}M_\mathrm{t} \overline{v^{2}_\mathrm{p}})$ is about
$ 1.8\times 10^{57}\,\mbox{ergs}$, which implies processes that generated at least a total of $ 10^{58}\,\mbox{ergs}$.
  This huge amount of energy should have been supplied by supernovae and  winds from massive stars 
in a starburst originated by  the collision of the Clouds. The collective action  of supernovae and stellar winds of 
a typical starburst  can drive a ``superwind'' that may eject  $10^{7}-10^{9}\, \mathrm{M}_{\sun}$ from a galaxy 
 with a mechanical energy of 
$10^{57}-10^{59}\,\mbox{ergs}$ during  its  lifetime estimated at  $10^{7}\,\mathrm{yr}$ (Heckman et al. \cite{heckman}). There 
are various stellar generations in the Magellanic Clouds 
(Westerlund \cite{westerlund}). There is nothing against the idea  that the formation of the younger generations, 
with ages lower than  600 Myr, was initiated by the powerful  starburst associated with the collision of the
 Clouds about 570 Myr ago.

\section{Interaction of the high velocity clouds with the Galactic disk and generation of the Galactic warp}
Most of the HVCs have not been affected by the friction force of the interstellar medium, since their orbits
have  not intersected with  the gas layer of the Galactic disk. Here,  we analyze the important group of  HVCs that was 
strongly decelerated by the friction force of the Galactic layer and that had a strong influence 
on the Galactic disk.      

The HVCs that penetrated the dense regions of the gaseous disk  were  trapped  in the Galactic plane, and are  participating in 
the Galactic rotation now (see  Figs. ~\ref{impactdisk1} and   ~\ref{impactdisk2}). 
\setcounter{newctr}{1}
\renewcommand{\thenewctr}{\alph{newctr}}
\renewcommand{\thefigure}{\arabic{figure}-\thenewctr}
 \begin{figure}
   \centering
 \includegraphics[width=\textwidth]{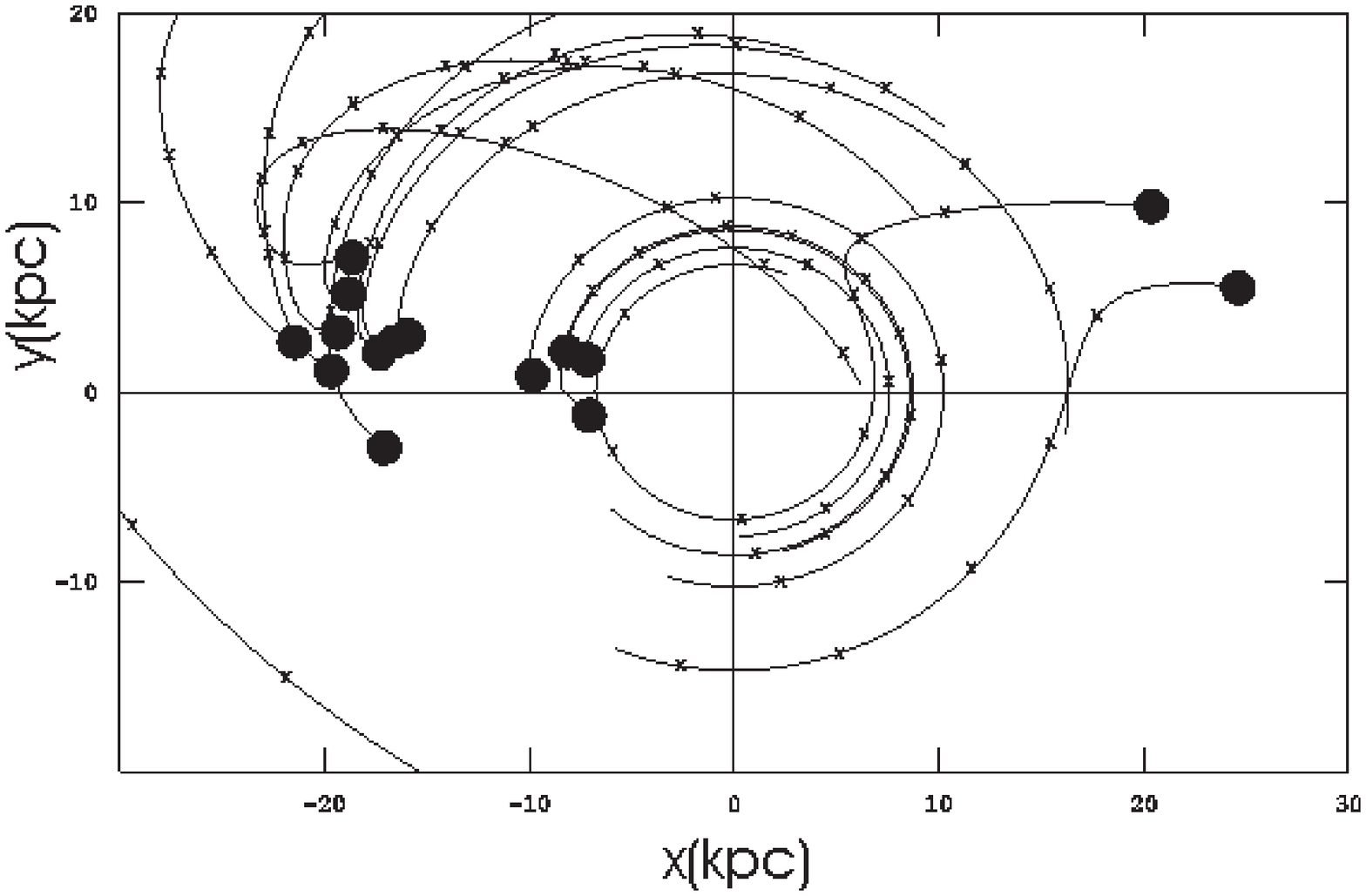}
      \caption{Projection, on the Galactocentric $X$-$Y$ plane,  of the orbits of some of the test particles that interact  with the Galactic disk. The filled circles indicate the positions of entry into the Galactic layer. The crosses mark the positions of the particle  at intervals of 30 Myr. The times of entry of this subset of  test particles lie  between -220 and  -120 Myr. The position of the Sun $(X_{\sun},Y_{\sun})$ is (0, 8.5)kpc. 
              }
         \label{impactdisk1}
   \end{figure}
\addtocounter{figure}{-1}
\addtocounter{newctr}{1}
\begin{figure}
   \centering
 \includegraphics[width=\textwidth]{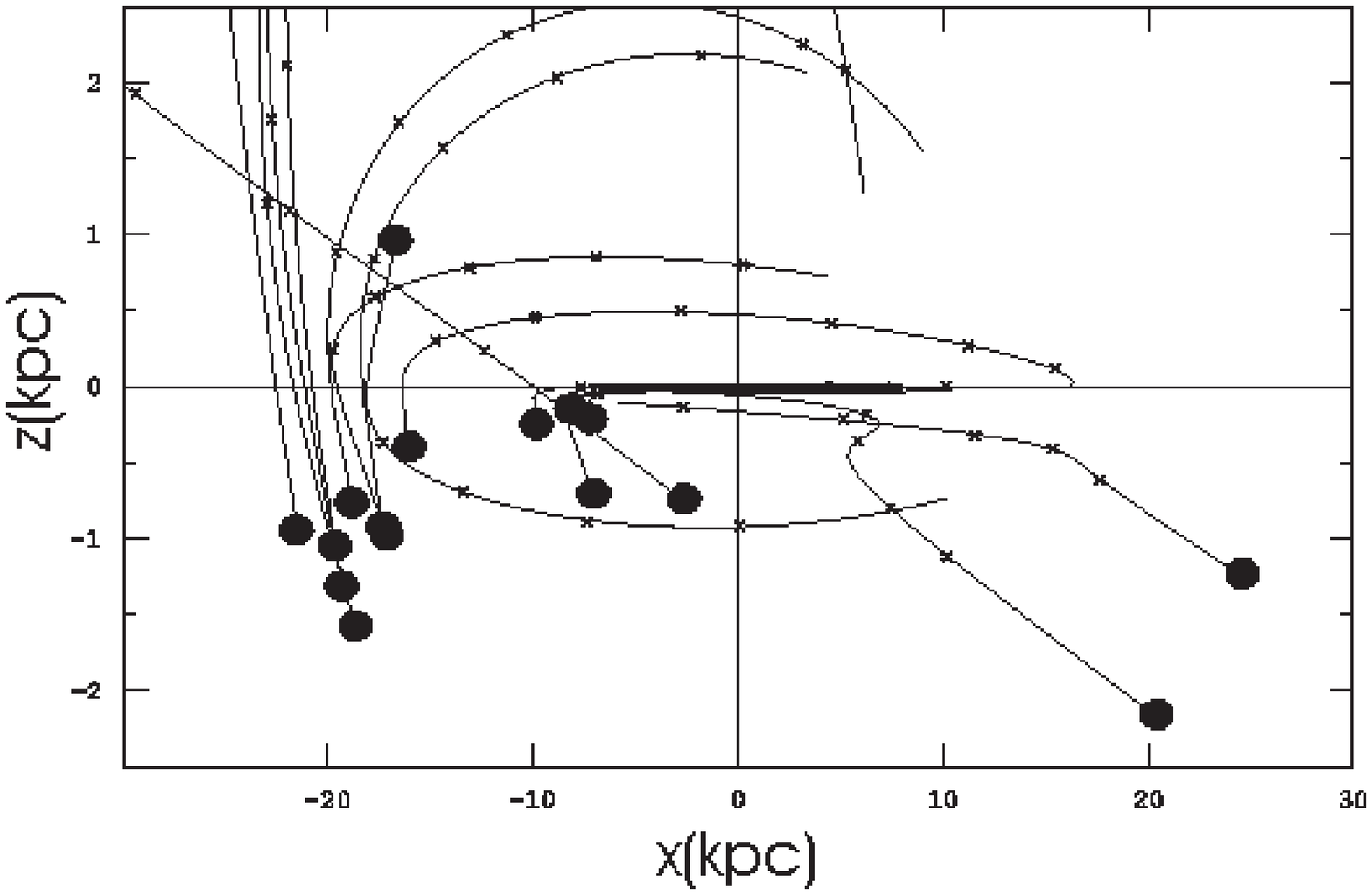}
      \caption{ Projection of the orbits of the same test particles as those of  Fig. ~\ref{impactdisk1}, but on the Galactocentric  $X$-$Z$ plane.
              }
         \label{impactdisk2}
   \end{figure}
In Fig.  ~\ref{impacts}, we show the positions of entry  of the test particles
in the gaseous layer of the Galactic disk and their present positions, resulting  from  the braking process. According 
to our model, during the last 400 Myr  the Galactic disk has been profusely hit by HVCs (see Fig. ~\ref{impacts} and Fig. ~\ref{fronta1}). 
\renewcommand{\thefigure}{\arabic{figure}}
\begin{figure}
 \includegraphics[width=\textwidth]{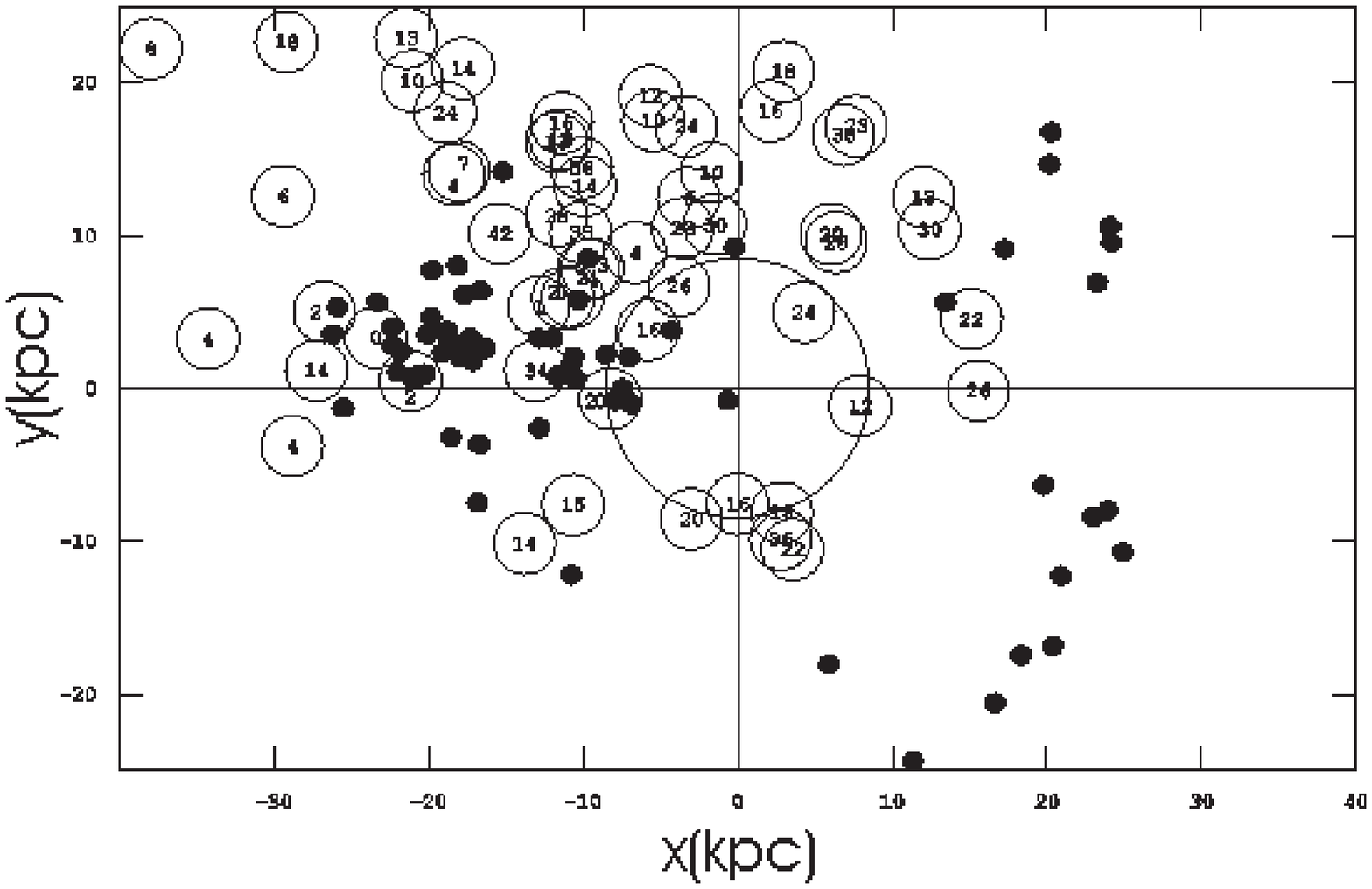}
      \caption{Same as Fig. ~\ref{impactdisk1}, but for the total set of test particles that interact with the Galactic layer. The filled and open  circles indicate respectively  the positions of entry into the Galactic  gas layer,  and  the present positions of the impacted regions,  resulting from the interaction of the test particles with a rotating disk. We have omitted the projection of their complete orbits for clarity. The numbers inside the circles refer to  the ages of impacted regions  in units  of 10 Myr, defining age as  the time  elapsed from the entry of the particle  into the Galactic layer up to the present.  The solar circle is indicated.
              }
         \label{impacts}
   \end{figure}
\setcounter{newctr}{1}
\renewcommand{\thenewctr}{\alph{newctr}}
\renewcommand{\thefigure}{\arabic{figure}-\thenewctr}
 \begin{figure}
   \centering
  \includegraphics[width=\textwidth]{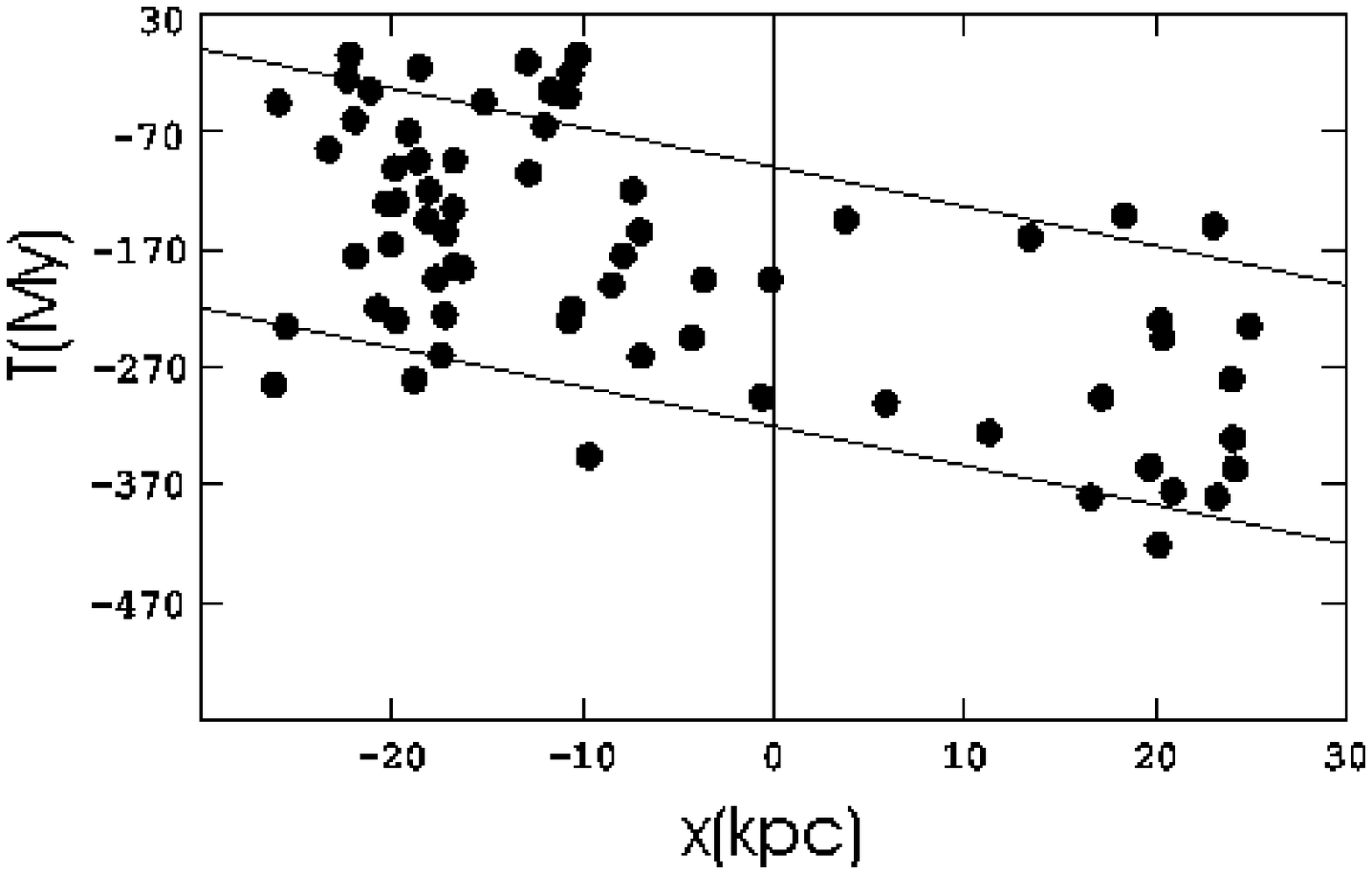}
      \caption{Correlation between the times of entry of the test particles into the Galactic layer and  the positions of the particles  projected on the $X$-axis. The solid lines represent  the position  of the front edge (lower line) and the back edge of the accretion front as a function of time. Negative time means before the present.  
              }
         \label{fronta1}
   \end{figure}
\addtocounter{figure}{-1}
\addtocounter{newctr}{1}
\begin{figure}
   \centering
  \includegraphics[width=\textwidth]{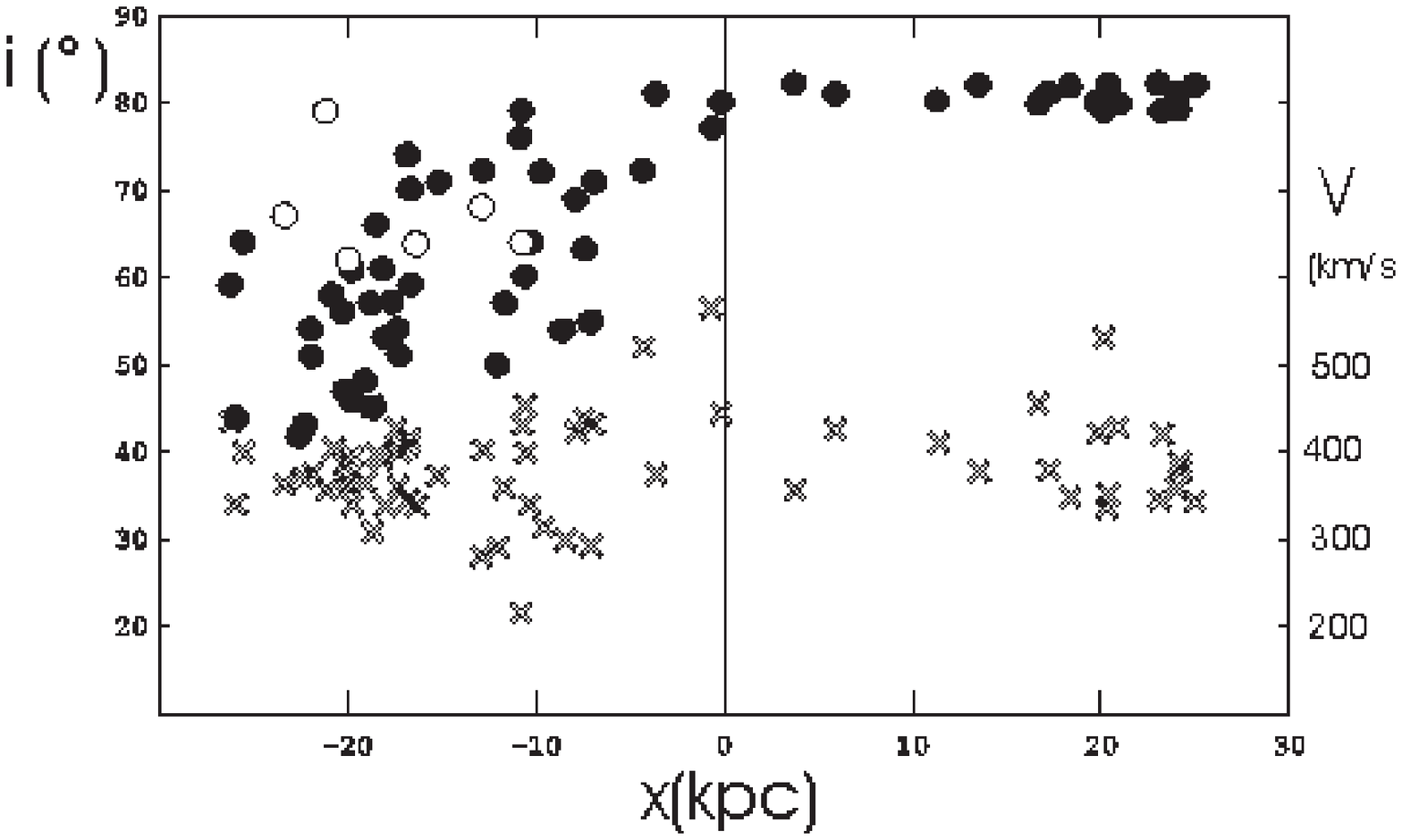}
      \caption{Angles (filled and open circles) and velocities (crosses)  of entry of the test particles into the Galactic layer versus the respective positions projected on the $X$-axis. The open circles refer to angles of entry from above the Galactic plane.
              }
         \label{fronta2}
   \end{figure}

The detailed description of the interaction of HVCs with the Galactic disk is beyond the scope of the present study. 
Tenorio-Tagle (\cite{tenorioa}, \cite{tenoriob}) analyzed the physics of the collisions of HVCs 
with the Galactic disk. It is clear that the impact of HVCs upon the Galactic disk should be an important source for 
producing shells and supershells. This mechanism of creation of large-scale structures in the interstellar medium
can account for many of the observational details of the large HI shells lying beyond the solar circle (Mirabel \cite{mirabelb}; Heiles \cite{heiles}). In connection with this, it is interesting to compare the theoretical  distribution of the impacts of test particles on the Galactic disk (Fig.~\ref{impacts}), with the location of observed  shells and supershells of HI (McClure-Griffiths et al. \cite{mcclure}).
 Evidence for interactions  of HVCs with the gas in the Galactic plane was found by several authors 
(see Tripp et al. \cite{tripp}  and the reviews of  Tenorio-Tagle \&
Bodenheimer \cite{tenorio-bodenheimer} and P\"{o}ppel \cite{poeppel}). 

 In the next paragraphs of this section we will show   that the flow  of  HVCs onto  the  Galactic disk  over a period of  400 Myr has caused a major 
disturbance in the Galaxy. As  a consequence of the fall of clouds, a burst of star formation involving   the whole Galactic disk  should have started  400 Myr ago and is still going on (Barry \cite{barry}; Noh \& Scalo \cite{noh}).  There are observational details of the vertical structure of the Galactic disk, such as  peculiar departures from 
the midplane or  huge depressions delineated by young star-gas supercomplexes (Alfaro et al. \cite{alfaro}), that may be explained in terms of 
 HVC-disk collisions that induced star formation
(Cabrera-Ca$\tilde{\rm n}$o et al. \cite{cabrera}). Further predictions provided by our model are the times, velocities and  angles of entry
  of the HVCs in the Galactic layer (see Figs.  ~\ref{fronta1} and  ~\ref{fronta2}). The angle of entry (or incidence) of an HVC is  defined as the angle
 of the direction of motion of an  HVC with respect to the vertical ($Z$-axis). The angle of entry  is an important parameter for the hydrodynamic simulations of the collision process (Comer\'{o}n \& Torra \cite{comeron};
 Franco et al. \cite{franco}). The times, velocities and angles of entry correspond  to the point at  which the 
friction coefficient reaches  the threshold value $\varepsilon\frac{n S}{m} = 10^{-5}\,\mathrm{pc^{-1}}$ (see Sect. 3.1).   Figs  ~\ref{fronta1} and  ~\ref{fronta2} show interesting differences  between the times and angles of entry on the right side of the Galactic disk  (positive $X$-axis) and those on the left side. The HVCs arrived first at the right side of the Galactic disk, while there was  a delay of $\sim 150\,\mathrm{Myr}$ for those arriving at  the left  side of the disk. On the right side of the disk the HVCs impinged on  the disk from below with angles of entry $\sim 80\degr$, i.e. almost a grazing incidence. In contrast, on the left side of disk the angles of entry are lower;  and although most of the HVCs entered  from below too, there were  some  HVCs that entered from above.  

In the following we examine whether there is a causal relationship  between the  HVC-disk interaction and
the generation of the Galactic warp. Galactic warps are very complex phenomena (see  L\'{o}pez-Corredoira et al.  \cite{lopez}, 
Garc\'{\i}a-Ruiz et al. \cite{garcia} and references therein). We will analyze  only a few
aspects of the problem in the light of our model. During the last  450 Myr, the flow of HVCs has been exerting  a pressure  on 
the gaseous disk which could have caused the distortion of the disk of the Galaxy. Since about  7  per cent  of the test particles of the simulation 
 impacts  within the galactocentric radius $R=$ 25 kpc, the mass in the form of HVCs accumulated  by the disk is  $0.07\times M_\mathrm{t}\sim 2.2\, 10^{8}\,\mathrm{M}_{\sun}$. Therefore, the rate of mass accretion per surface unit is   $\gamma = 600\,\mathrm{M}_{\sun}\,\mathrm{kpc^{-2}\, Myr^{-1}} $, taking into account that the time during  which the flow acted on the disk at a fixed point in the $(X,Y,Z)$ frame  was  nearly  200 Myr (see Fig.~\ref{fronta1}). Let us  consider a cylinder of unit cross-section with generators parallel to the $Z$-axis and height $h(R)$ as  a volume element of the gaseous disk of the Galaxy at $R$. The accretion process acting on the volume element generates  a 
force in $Z$ per unit of mass  that   can be written  as
   $F_{z}=\frac{\gamma}{\sigma(R)}  \overline{v} cos(\overline{i})$, where $\overline{v}$, $\overline{i}$ are the mean velocity and angle of entry, and $\sigma (R)$ is the surface density of  the gaseous disk at $R$ (see Sect. 3.1).

To describe  the gas  motion of the volume element under the effects of the additional force associated with the mass accretion, we  use  cylindrical coordinates, $(R,\theta, Z)$, representing the galactocentric radius, azimuth and  vertical distance from    the plane $b=0\degr$ of the mass center of the chosen volume element.  
The gas motion in the $Z$-direction is governed by  the force $F_{z}$ and the restoring force of the stellar disk $F=-\lambda^{2}Z$. It can be expressed by \begin{equation}Z(t)=Z_{0}(t_\mathrm{e})+ \frac{sin(\lambda (t-t_\mathrm{e}))}{\lambda} \int_{0}^{t-t_\mathrm{e}} F_{z} cos(\lambda\tau) d\tau-
  \frac{cos(\lambda (t-t_\mathrm{e}))}{\lambda} \int_{0}^{t-t_\mathrm{e}} F_{z} sin(\lambda\tau) d\tau \end{equation}, (see Eq. 5.713 of
Chandrasekhar \cite{chandrasekhar}). As  initial conditions of the disk in $Z$ we put  $Z_{0}=\dot{Z}_{0}=0$ at the time origin $t_\mathrm{e}=-570\,\mbox{Myr}$. Ignoring the effects of the $(X,Y)$ component of the  force associated with the mass accretion,  and assuming circular rotation, we have $\theta(t)=\theta_{0}(t_\mathrm{e})+\frac{V_{c}}{R}(t-t_\mathrm{e})$. 

The flow of  HVCs moves parallel  to the $X$-axis. Therefore, we can think  of the interaction of the HVCs with the disk as 
an accretion front perpendicular to the $X$-axis,  propagating in the $X$ direction with a constant velocity $v_\mathrm{a}$. Then the position of the  accretion front as a function of time is $X_\mathrm{a}(t)=v_\mathrm{a}t +X_{0}(t_\mathrm{e})$ (see Fig.~\ref{fronta1}). The volume element  enters the front when
 $X =R\,  sin(\theta(t))=X(t)_\mathrm{a}$, a condition that allows us to determine the entry time $t_{1}$. The time at  which the volume element  leaves the rear zone of the accretion front,  $t_{2}$,  is determined by  solving  $X=X(t)_\mathrm{a}+ \Delta X_\mathrm{a}$ for t, where $\Delta X_\mathrm{a}$ is the thickness of the accretion front. The times $t_{1}$ and $t_{2}$ and the duration of the process of mass accretion, $ t_\mathrm{a}= t_{2}-t_{1}$,  depend  on   $\theta$ and $R$ of the volume element  (see Fig. ~\ref{warp1}). At large galactocentric radii the time intervals  of mass accretion $t_\mathrm{a}$ in  regions of the third and fourth Galactic quadrants are longer than those in   the first and second quadrants, i.e  the outer regions in  the third and fourth Galactic quadrants   are the most affected by the interaction with the HVCs. For each volume element, the force $F_{z}$ acts  between the corresponding times  $t_{1}$ and  $t_{2}$,  vanishing outside this time interval.  Hence,  Eq.~(11) becomes $Z(\theta, R, t)=\frac{F_{z}}{\lambda^{2}}(cos\lambda(t_{2}+ t_\mathrm{e}-t) - cos\lambda(t_{1}+t_\mathrm{e}-t))$. This is  a general expression for the $Z$ deformation of the Galactic layer due to the flow of  HVCs. At t=0 this equation can reproduce roughly  the present 
configuration of the Galactic warp, if the period of vertical oscillation $\frac{1}{2\pi\lambda}\sim 450\,\mathrm{Myr}$ (see Fig. ~\ref{warp2}). This  value seems  reasonable, because at large galactocentric radii the restoring force  should be small. Note that the model can explain the overall pattern of  positive vertical displacements  in the first and second Galactic quadrants, as well as the less pronounced  displacements toward  negative $Z$ in the third and fourth  quadrants (cf. our Fig.~\ref{warp2} with Figs. 57b and 60b of Burton \cite{burton}).    

The first explanation proposed for   the Galactic warp was that the Galaxy is passing through a continuous intergalactic medium (Kahn \& Woltjer \cite{kahn}). In contrast, we propose an interaction of the Galactic disk with a transient circumgalactic flow of  HVCs produced by a single event  of mass transfer from  the Magellanic Clouds. Our model shows that the flow of  HVCs is able to produce the formation of the Galactic warp. Future extensions of the model should  contemplate consequences of the process of mass accretion  such as large deviations from the circular rotation of the gaseous disk and   the sudden increment of the rate of  star formation and  of the input of  energy in the interstellar medium. The HVCs have injected   energy at a rate of $\sim 10^{50}\,\mathrm{ergs\, kpc^{-2}\,Myr^{-1}}$ into the Galactic disk  during the last 400 Myr. A question we should address, among others, should be  the role  played by this important source of energy in the production and maintenance of the diverse phases  of the interstellar medium  (Kulkarni \& Heiles \cite{kulkarni}; Elmegreen \cite{elmegreen}; Heiles \& Troland \cite{heiles-troland}). 
Another interesting question is whether there exists  any genetic or dynamic link between the Galactic warp and a ring of stars in the plane of the Galaxy, at a Galactocentric radius of 18-20 kpc, which may completely encircle the Galaxy (Ibata et al. \cite{ibata}; Yanny et al. \cite{yanny}). 
\setcounter{newctr}{1}
\renewcommand{\thenewctr}{\alph{newctr}}
\renewcommand{\thefigure}{\arabic{figure}-\thenewctr}
 \begin{figure}
   \centering
  \includegraphics[width=\textwidth]{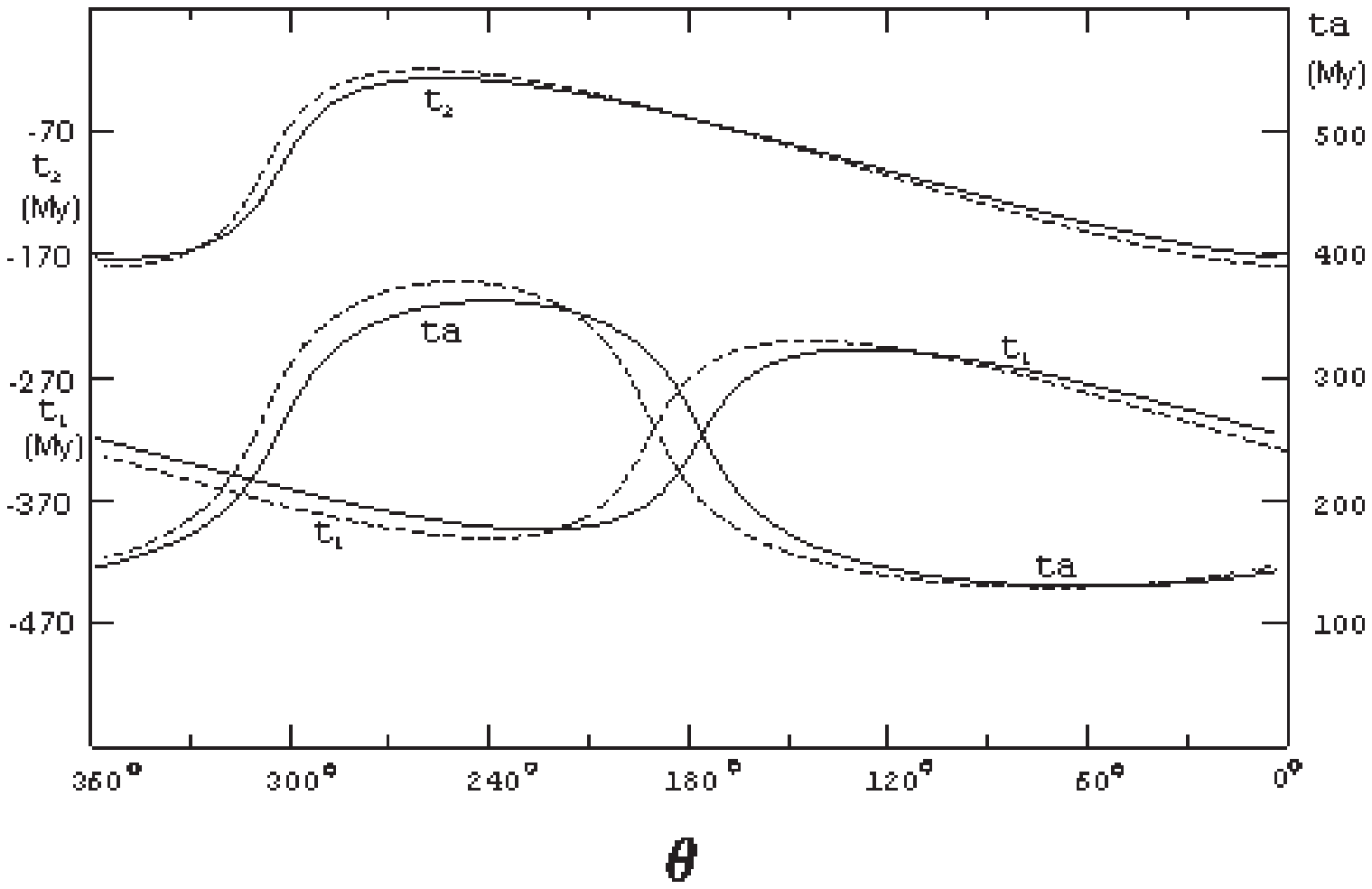}
      \caption{ The time  $t_{1}$ at which the head of the accretion front crosses a disk  region, and  the time  $t_{2}$ at which the back of the accretion front leaves the region,  as a function of the present azimuth for  the regions at $R=$ 22 kpc (dashed line) and  $R=$ 24 kpc (solid line). The duration of the transit of the accretion front, $t_\mathrm{a}=t_{2}-t_{1}$, 
 is also represented as a function of the present azimuth.}
         \label{warp1}
   \end{figure}
\addtocounter{figure}{-1}
\addtocounter{newctr}{1}
\begin{figure}
   \centering
  \includegraphics[width=\textwidth]{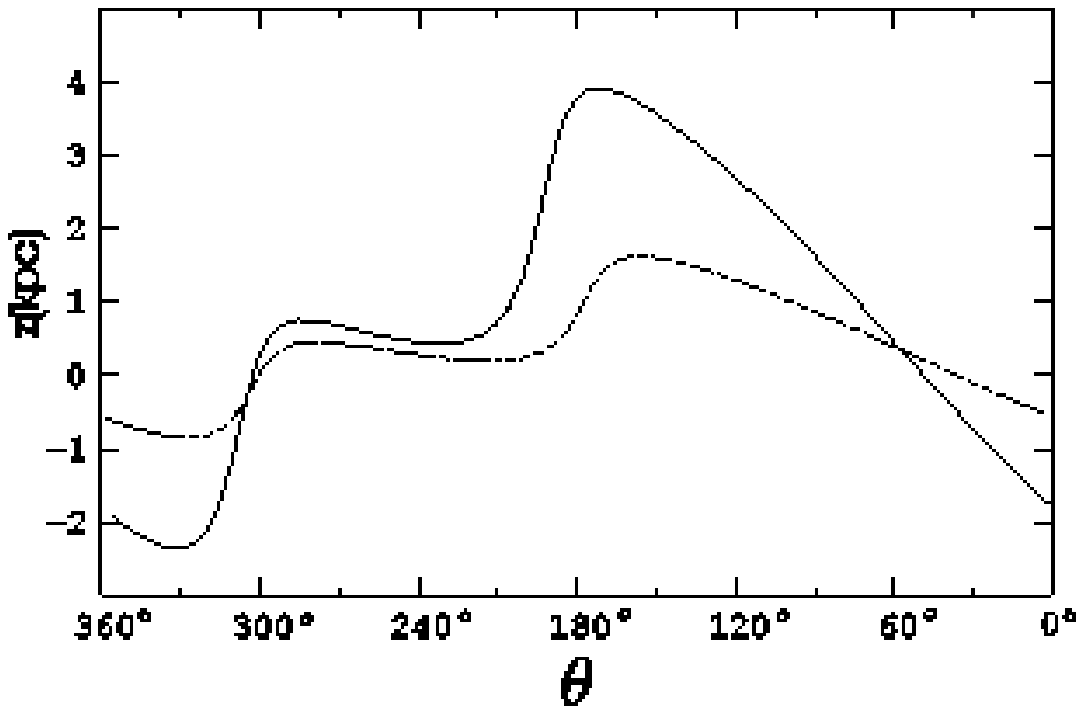}
      \caption{ Vertical $Z-$distances of  the midplane of the Galactic layer from the plane $b=0$  at the Galactocentric 
distances $R=$ 22 kpc (dashed line) and $R=$ 24 kpc (solid line)  as a function of  the azimuth of the region, according to the simulation of the  interaction between the accretion front due to the flow of HVCs and the Galactic gas layer. 
              }
         \label{warp2}
   \end{figure}

\section{Conclusions} 
We have shown that the majority of the HVCs have  a common  origin  linked to the dynamic evolution of
the Magellanic Clouds. During  a long time these two small satellite galaxies have likely  behaved as a binary system 
under the effects of the tidal perturbations of the Galaxy. The two Clouds have probably approached each other 
several 
times in the past, until the Clouds collided 
catastrophically  about 570 Myr ago.  We assumed that the collision triggered a burst of stellar formation 
in the Clouds. The superwind driven by the starburst blew the HVCs as magnetized clouds into 
the massive halo of the Galaxy. As a  consequence, the Milky Way is likely surrounded by a mist of highly ionized gas at  high velocity,  the HVCs being the denser  concentrations.

 The catastrophic loss of matter was of  a significant amount. Probably it  
permitted the SMC to reach escape velocity,
 disrupting the binary system. According to  the results of our model the amounts of matter and mechanical energy 
liberated in the form of HVCs were
 of the order of  $ 3.1 \times 10^{9}\,\mathrm{M}_{\sun}$ and $1.8\times 10^{57}\,\mathrm{ergs}$, respectively.  In the numerical simulation of the process, we assumed that
the HVCs were ejected from  the Clouds  570 Myr ago in all directions with  peculiar velocities 
larger  than the escape velocity 
of $160\,\mathrm{km\, s^{-1}}$,  and following  a Gaussian distribution law. Because of these initial expansion velocities,
the  cloud of HVCs, or metacloud,  expanded from the center of collision of the Magellanic Clouds,  reaching  its
present size of $\sim 300\,\mbox{kpc}$, and  filling  a great part of the halo volume  of the Galaxy.
As the initial mean velocity of the HVCs was  that of the Clouds at the time of the ejections of  HVCs,
 the centroid of the group of HVCs (metacloud) describes an  orbit close to the orbit of  the Magellanic Clouds 
(i.e.  approximately  in the direction $\ell=90\degr-270\degr$). To an observer at the  position of the Sun, this gives  
the impression of a flow of HVCs coming from $\ell=90\degr$. As a consequence of the initial velocity distribution of 
the HVCs, 
a larger number of HVCs tended   to  concentrate towards the Magellanic orbit behind the present position of the Clouds, 
whereas other HVCs moved ahead of  them, forming the Magellanic Stream.   

The passage of the HVC flow through the Galactic disk had transcendent consequences for the evolution of the gaseous layer of the Galaxy. The outer gas layer of the Galaxy was considerably displaced in the $Z$-direction from its equilibrium position. The Galactic disk accumulated mass coming in the form of HVCs  at an average rate of  $600\,\mathrm{M}_{\sun}\,\mathrm{kpc^{-2}\, Myr^{-1}}$ over a period of 200 Myr.  Due to the time dependence  of  the mass accretion upon  the azimuth of the region, the effects on the first and second Galactic quadrants at large galactocentric radii were different from those on the third and fourth ones. These are perhaps some of the clues to understand the genesis of the Galactic warp. The collisions of the HVCs with the Galactic disk  might have exerted notable influences on the large-scale morphology and energy  of the interstellar medium.
     
%
  

\begin{acknowledgements}
I am particularly grateful to Dr. Virpi  S. Niemela and Dr. Wolfgang G. L. P\"{o}ppel for their  constant encouragement and  help. Dr. Wolfgang G. L. P\"{o}ppel read the manuscript with a constructive eye. I am indebted  to Dr. Bart P. Wakker for providing his catalog of HVCs via e-mail. The helpful comments of an anonymous referee led to substantial  improvements in the  paper.   Part of this work was supported by the \emph{Consejo Nacional de Investigaciones Cient\'{\i}ficas  y T\'{e}cnicas (CONICET)} 
       project
      number PIP-0608/98. 
\end{acknowledgements}

\end{document}